\documentclass[pdflatex,sn-basic]{sn-jnl}

\usepackage[utf8]{inputenc}
\usepackage{graphicx}
\usepackage{multirow}
\usepackage{amsmath,amssymb,amsfonts}
\usepackage{amsthm}
\usepackage{mathrsfs}
\usepackage[title]{appendix}
\usepackage{xcolor}
\usepackage{textcomp}
\usepackage{manyfoot}
\usepackage{booktabs}
\usepackage{algorithm}
\usepackage{algorithmicx}
\usepackage{algpseudocode}
\usepackage{listings}

\usepackage{caption}
\usepackage{placeins}

\usepackage{hyperref}

\theoremstyle{thmstyleone}

\theoremstyle{thmstyletwo}

\theoremstyle{thmstylethree}

\begin{document}

\title[
  Modelling Territorial Dynamics through Statistical Mechanics:
  An Application to Resident Foreign Population
]{
  Modelling Territorial Dynamics through Statistical Mechanics:
  An Application to Resident Foreign Population
}

\author*[1]{\fnm{Pierpaolo} \sur{Massoli}}
\email{pimassol@istat.it}

\affil[1]{%
  \orgdiv{Directorate for Methodology and Statistical
    Process Design (DCME)},
  \orgname{Italian National Institute of Statistics (ISTAT)},
  \orgaddress{%
    \street{Via Cesare Balbo 16},
    \city{Rome},
    \postcode{00184},
    \country{Italy}
  }
}


\abstract{
This study proposes an energy-based framework for exploring
local variations around observed territorial configurations
in Official Statistics. The observed distribution serves as
the initial reference, without assuming equilibrium.
A register-derived graph represents structural similarities
between municipalities, while socio-economic composite indices
are synthesised through Principal Component Analysis into
a univariate external field. 
A Continuous Ising model and a Langevin-based procedure
provide complementary stochastic exploration strategies.
Simulated Annealing progressively limits exploration,
favouring lower-energy configurations under each specification.
Recorded energy trajectories and municipal deviations
characterise the resulting ensembles. Conformal Prediction
provides calibrated intervals whose widths and empirical
coverage describe their relationship with fixed observed
municipal values under repeated subsampling.

The application to resident foreign population shares in
North-East Italy yields simulated municipal averages close
to the observed values, but substantially different calibrated
interval widths between the two procedures. Comparisons across
territorial groups reveal heterogeneous associations with
socio-economic profiles, without a uniform ordering of interval
widths between central and peripheral municipalities. 
The framework connects territorial interactions, socio-economic
information, and stochastic variation within an exploratory
analysis. Its contribution is to examine how nearby simulated
alternatives relate to the observed configuration and to
formulate hypotheses about territorial heterogeneity.
Further applications and sensitivity analyses would assess
the robustness of these interpretations.
}





\maketitle


\section{Introduction}\label{sec1}

Geo-referenced demographic and economic data provide
opportunities for analysing territorial systems in Official
Statistics. Understanding socio-demographic distributions
requires consideration of both local characteristics and
relationships among territorial units. An observed configuration
can therefore be studied as the state of an interacting system,
whose organisation reflects demographic, economic, and
territorial factors. This perspective motivates an examination
of how the system responds to specified interactions and
socio-economic drivers from its observed starting condition. 
Established approaches address complementary aspects of
territorial analysis. Spatial econometrics models regional
dependence through spatial autocorrelation
\citep{anselin1988,bivand2013}, although specifications based
on linearity and normality may inadequately represent complex
non-linear relationships. Geographical proximity also need
not imply socio-economic similarity: distant municipalities
may share structural characteristics, while neighbouring
ones differ substantially. Relationships defined exclusively
by distance may consequently overlook relevant connections. 
Composite indicators synthesise multidimensional information
for comparison and policy evaluation
\citep{istat2024,noll2004social,greco2019mcda}. Additive
aggregation permits compensation among dimensions, while
non-compensatory alternatives \citep{mazziotta2016} limit
such trade-offs. These summaries primarily characterise
individual units; representing their interactions requires
an additional relational structure. Machine learning captures
complex associations in high-dimensional data, including
poverty mapping with convolutional neural networks
\citep{jean2016} and urban-peripheral classification with
boosted trees \citep{casacci2024}. These applications emphasise
prediction and classification rather than the response of an
observed territorial configuration to specified forces. 
Graph-based representations connect units through attribute
similarities or network relationships beyond geographical
proximity \citep{grubesic2026}. Agent-based and cellular
models likewise provide established representations for
simulating urban dynamics \citep{batty2021simulating}.
Statistical Mechanics offers a complementary framework for
examining interacting systems, with applications to collective
decision making, wealth distribution, and network diffusion
\citep{galam1997,durlauf1999,bouchaud2000,castellano2009,
Lelarge2015}. In the present application, energy is a
model-defined criterion for evaluating territorial
configurations under internal interactions and external
forces, rather than a physical quantity. 
The proposed framework addresses a substantive question:
to what extent is the observed territorial configuration
compatible with interactions and socio-economic drivers
specified from statistical registers? Observed municipal
values define the initial state and fixed empirical reference,
without being assumed to represent an equilibrium or local
energy minimum. Simulated configurations evolve while the
interaction graph and external field remain unchanged.
Their deviations from the reference provide an exploratory
basis for examining territorial differences under these
modelling choices. 
Two stochastic strategies implement this perspective.
A continuous (soft-spin) formulation extends the binary
Ising model \citep{ising1925}, representing municipalities
through continuous states and graph-based couplings.
Composite socio-economic indices are synthesised through
Principal Component Analysis into a univariate external
field \citep{Pareto2025}. A complementary Langevin-based
procedure combines gradient-driven updates with random
perturbations \citep{Kloeden1992,Risken1996,Schlick2002,
Welling2011}. Following stochastic relaxation principles
\citep{geman1984}, both procedures explore alternatives
from the same observed state, using distinct state scaling
and transition rules. 
Monte Carlo simulation is combined with Simulated Annealing
\citep{metropolis1953,aarts1989,binder1997} and parallel
computing \citep{mccallum2011,weston2017}. Cooling progressively
limits stochastic exploration; the objective is to examine
nearby, energetically favourable configurations rather than
find a global optimum. Proximity is assessed empirically
through municipal deviations. These local variations, or
``vibrations'', are related to socio-economic profiles and
territorial attributes. 
Conformal Prediction
\citep{vovk2005,shafer2008,burnaev2014,lei2018,
angelopoulos2021,tibshirani2023,unece2026uq}
complements the analysis through calibrated interval widths
and empirical coverage of fixed observed municipal values
under repeated subsampling. Their spatial representation
describes territorial differences under the simulation and
calibration protocol. The application to resident foreign
population shares in North-East Italy illustrates how this
framework can support Official Statistics in exploring
territorial organisation and formulating hypotheses about
its association with register-derived socio-economic
characteristics.


\section{Theoretical background}\label{sec2}

This section introduces the Continuous Ising model, Langevin
dynamics, composite indices, and Conformal Prediction.

\subsection{The Continuous Ising model}
\label{sec2_1}

Originally developed to describe ferromagnetism, the Ising model
represents interacting systems through network nodes with binary
states $s_i \in \{-1,+1\}$, governed by interactions and external
fields. Its continuous \textit{soft--spin} formulation instead
allows $s_i \in \mathbb{R}$ to represent real--valued indicators,
such as municipal foreign resident percentages, with the
admissible state space restricted to their support.
For a configuration $\mathbf{s} = (s_1,\ldots,s_N)$,
a quadratic Hamiltonian is
\begin{equation}
H(\mathbf{s}) = -\tfrac{1}{2}\sum_{i,j} J_{ij} s_i s_j
       - \sum_i h_i s_i
       + \tfrac{\lambda}{2}\sum_i s_i^2
\label{eq:hamilt}
\end{equation}
where $J_{ij}$ encodes pairwise interactions, $h_i$ is the
external field, and $\lambda>0$ penalises large spin values.
For symmetric $J$ on $\mathbb{R}^N$, normalisability requires
$\lambda I-J$ to be positive definite; $\lambda>0$ alone
is insufficient. On a compact state space, a continuous
Hamiltonian has a finite partition function at any fixed
positive temperature.
At temperature $T>0$, the configuration density is
\begin{equation}
P(\mathbf{s}) = \frac{1}{Z}\exp\!\left(-\frac{H(\mathbf{s})}{T}\right)
\label{eq:boltzmann}
\end{equation}
with partition function over the admissible state space
$\mathbb{S}$
\begin{equation}\label{eq:partition}
Z = \int_{\mathbb{S}}
    \exp\!\left(-\frac{H(\mathbf{s})}{T}\right)
    \,d\mathbf{s}.
\end{equation}
For discrete states, the integral becomes a sum.
Throughout this paper, configurations are compared through
relative Gibbs densities at a common temperature within each
energy specification. Their logarithmic ratio is termed the
\textit{log-likelihood ratio}, exclusively in this energy-based
sense, rather than as a likelihood for classical inference.

Temperature controls the weighting of energy differences,
with lower energies favoured. The positive-definite quadratic
case on $\mathbb{R}^N$ admits a Gaussian closed form for $Z$;
other energy functions or state-space restrictions may
complicate its evaluation. Monte Carlo simulation explores
configurations without computing $Z$ explicitly.
At iteration $t$, a randomly selected node receives a proposal
$s_i^{\ast} = s_i + \eta$, typically with Gaussian $\eta$.
For a symmetric proposal mechanism on the admissible state
space, acceptance occurs with probability
\begin{equation}
\alpha = \min\!\left\{
  1, \exp\!\left(-\frac{\Delta H}{T(t)}\right)
  \right\},
\label{eq:acceptance_cont}
\end{equation}
where $\Delta H$ is the induced energy change.
At fixed temperature, this rule satisfies detailed balance
with the Gibbs distribution; convergence additionally requires
suitable conditions on the transition mechanism. 
Simulated Annealing decreases $T(t)$, progressively reducing
acceptance of energy-increasing moves that, at higher
temperatures, can facilitate escape from shallow local minima.
The resulting time-inhomogeneous process generally has no
single stationary Gibbs distribution throughout cooling.
Exploration depends jointly on the proposal mechanism,
cooling schedule, and simulation duration.

\subsection{Langevin dynamics}\label{sec2_2}

Introduced for Brownian motion, Langevin dynamics describes
continuous--time trajectories under deterministic forces and
random perturbations. For a particle with position $x(t)$
and velocity $v(t)$, the classical equation is
\begin{equation}
m \frac{d v(t)}{dt} = -\gamma v(t) + F(x(t))
  + \sqrt{2 \gamma T}\,\xi(t),
\label{eq:langevin_classic}
\end{equation}
where $m$ is mass, $\gamma$ friction, $F(x)$ the force field,
and $T$ temperature, with the Boltzmann constant set to one.
Gaussian white noise $\xi(t)$ has zero mean and covariance
$\mathbb{E}[\xi(t)\xi(t')]=\delta(t-t')$.
Neglecting inertia, taking the force as the negative energy
gradient, and choosing unit mobility gives the overdamped form
\begin{equation}
\frac{d s_i(t)}{dt} = -\frac{\partial H(\mathbf{s})}{\partial s_i}
  + \sqrt{2 T}\,\xi_i(t),
\label{eq:langevin_overdamped}
\end{equation}
where $s_i(t)$ is the state of node $i$, $H(\mathbf{s})$ the
Hamiltonian, and $\xi_i(t)$ independent Gaussian white noises.
The two terms represent gradient descent and stochastic
exploration, respectively.

The corresponding probability density $P(\mathbf{s},t)$
evolves according to the \emph{Fokker--Planck} equation:
\begin{equation}
\frac{\partial P(\mathbf{s},t)}{\partial t} = \nabla \cdot \left(
  \nabla H(\mathbf{s}) P(\mathbf{s},t) + T \nabla P(\mathbf{s},t)
  \right).
\label{eq:fokker_planck}
\end{equation}
At fixed temperature, the normalisable Boltzmann density is
a stationary solution under appropriate boundary conditions.
Reflecting boundaries supply a compatible zero-flux condition
on bounded domains. Convergence to equilibrium additionally
requires regularity and ergodicity conditions. 
The Euler--Maruyama approximation of
Equation~\ref{eq:langevin_overdamped}, with step $\Delta t$, is
\begin{equation}
s_i^{(t+1)} = s_i^{(t)} - \Delta t \,
  \frac{\partial H(\mathbf{s}^{(t)})}{\partial s_i}
  + \sqrt{2 T \Delta t}\,\eta_i^{(t)},
\label{eq:euler_maruyama}
\end{equation}
where $\eta_i^{(t)} \sim \mathcal{N}(0,1)$ are independent.
At finite step size, this scheme generally does not preserve
the exact Gibbs distribution. It has no acceptance step;
Metropolis-adjusted variants account for proposal asymmetry
through an appropriate acceptance rule.
For fixed $\Delta t$, cooling reduces noise but does not
directly reduce the gradient-driven drift. Decreasing the
step size can reduce both increments.

\subsection{Continuous Ising model vs. Langevin dynamics}
\label{sec2_3}

Metropolis simulation of the Continuous Ising model and
Langevin dynamics offer complementary exploration mechanisms:
the former accepts local proposals according to energy
differences (Equation~\ref{eq:acceptance_cont}), whereas the
latter combines gradient-driven motion with noise
(Equation~\ref{eq:euler_maruyama}). 
With the same Hamiltonian, state space, and fixed temperature,
a suitable Metropolis sampler and continuous--time Langevin
process can share an invariant Gibbs distribution. This
relationship does not automatically extend to finite-step
discretisations, different state scalings, or modified
acceptance rules. Their distinct responses to cooling,
described above, likewise do not establish equivalent
trajectories. 
Comparing the resulting ensemble summaries can therefore
identify common and differing empirical patterns without
requiring a common stationary distribution or assuming
that the initial reference configuration is an equilibrium.

\subsection{Methods for creating composite indices}
\label{sec2_4}

Composite indices synthesize multi-dimensional socio-economic
information into a single summary measure to facilitate comparisons
among statistical units. They are categorized into non-compensatory
methods, suitable when all input dimensions are essential, and
compensatory methods, which allow low values in one indicator to be
offset by high values in another.

\paragraph{A popular non-compensatory composite index}
A well-known non-compensatory method is the \textit{Mazziotta-Pareto
Index} (MPI), which standardizes base indicators into $z$-scores and
penalizes unbalanced unit profiles. Given an input dataset
$\mathbf{X}=\{x_{ij}\}$ of $j=1,\ldots,M$ indicators across
$i=1,\ldots,N$ statistical units, each variable is standardized as
\begin{equation}
x_{ij}^{s} = 10 \cdot pol_{j} \cdot \left(
  \frac{x_{ij} - \mu_j}{\sigma_j} \right) + 100
\end{equation}
where $pol_j \in \{-1,+1\}$ is the polarity, $\mu_j$ the mean, and
$\sigma_j$ the standard deviation of indicator $j$. 
The MPI is subsequently computed as
\begin{equation}
MPI_i = M_i \pm S_i \cdot \frac{S_i}{M_i},
\end{equation}
where $M_i$ and $S_i$ are the mean and standard deviation of the
standardized profile of unit $i$, respectively. The sign $\pm$
depends on the direction of the phenomenon under study. A higher
dispersion among the input indicators increases the penalty term,
reflecting penalization for unbalanced units.

\paragraph{An effective compensatory composite index}
\emph{Principal Component Analysis} (PCA) is a compensatory technique
that transforms correlated variables into uncorrelated principal
components $\mathbf{pc}_1, \mathbf{pc}_2, \ldots, \mathbf{pc}_M$. In
this framework, components are aggregated into a weighted composite
index $\mathbf{C}_\text{PCA}$ using the proportion of explained
variance as weights:
\begin{equation}
\mathbf{C}_\text{PCA} = \lambda_{1}\mathbf{pc}_{1}
  + \lambda_{2}\mathbf{pc}_{2} + \ldots + \lambda_{M}\mathbf{pc}_{M}
\end{equation}
where $\lambda_i$ indicates the $i$-th normalized eigenvalue
($\sum \lambda_i = 1$). This aggregation ensures that the summary
index reflects the dominant structure of variability in the original
data.

\subsection{Conformalized Quantile Regression approach}\label{sec2_5}

Conformal Prediction (CP) is a distribution--free and model--agnostic
framework for constructing prediction intervals with marginal
coverage guarantees under exchangeability. Combined with quantile
regression, \emph{Conformalized Quantile Regression} (CQR) produces
intervals whose widths adapt to heteroscedasticity without specifying
a parametric distribution for the response
\citep{lei2018, romano2019, angelopoulos2021}. 
Given a dataset of $N$ observations $\{(X_i, Y_i)\}_{i=1}^N$, CQR
constructs an interval $\mathcal{C}(X)=[L(X),U(X)]$ for a new
observation $(X,Y)$ such that
\begin{equation}\label{eq:CP1}
\Pr(Y \in \mathcal{C}(X)) \geq 1 - \alpha,
\end{equation}
where $\alpha \in (0,1)$ denotes the target miscoverage level.
This guarantee is marginal over the calibration data and the new
observation; it does not generally imply coverage at the nominal
level for every fixed value of $X$. 
In standard split CQR, the available observations are partitioned
into a training dataset and a separate calibration dataset.
The training dataset is used to estimate the lower and upper
conditional quantiles at levels $\alpha/2$ and $1-\alpha/2$.
These estimates are then held fixed when computing the
nonconformity score for each calibration point:
\begin{equation}\label{eq:CP2}
s(x, y)=\max\!\left[\hat{\tau}_{\alpha/2}(x)-y,\;
  y-\hat{\tau}_{1-\alpha/2}(x)\right].
\end{equation}
Let $n$ denote the number of calibration observations. The
adjustment $\hat{q}$ is the empirical quantile of the calibration
scores at level $\lceil (n+1)(1-\alpha)\rceil/n$, selected without
interpolation as the $\lceil (n+1)(1-\alpha)\rceil$-th smallest
score. If this rank exceeds $n$, the adjustment is taken to be
positive infinity. The conformalized prediction interval for a
new test point $X_{\text{test}}=x$ is constructed as
\begin{equation}\label{eq:CP3}
\mathcal{C}(x)= [\tilde{\tau}_{\alpha/2}(x)-\hat{q},\;
  \tilde{\tau}_{1-\alpha/2}(x)+\hat{q}],
\end{equation}
where $\tilde{\tau}$ denotes the same training-based quantile
estimates denoted by $\hat{\tau}$ in Equation~\ref{eq:CP2},
evaluated at the test point. No additional quantile model is
fitted during calibration. 
The score can take negative values when the response lies
strictly inside the initial interval. Consequently, the
calibration adjustment can contract as well as expand its
bounds. Under exchangeability of the calibration and test
observations, conditional on the training data, the rank-based
adjustment yields the marginal coverage guarantee in
Equation~\ref{eq:CP1}. The accuracy of the initial quantile
estimates affects interval efficiency, but the coverage
guarantee does not require a correctly specified quantile
regression model.


\section{Proposed approach}\label{sec3}

The proposed methodology investigates the compatibility of an
observed territorial configuration with a specified interaction
structure and socio-economic field. Municipalities are represented
as interacting particles with continuous state variables and
register-derived territorial attributes. The observed municipal
values define the initial configuration and remain fixed as the
empirical reference, while simulated states evolve under a constant
interaction graph and external field. The resulting trajectories
describe a simulated transient: simulation time indexes stochastic
exploration, not the historical development of the territory.
The analysis examines whether energetically more favourable
configurations preserve the observed territorial organisation and
how their deviations vary across municipal profiles.
\emph{Structural coherence} denotes this descriptive agreement,
assessed through municipal deviations and territorial comparisons. 
Two components specify the energetic evaluation of configurations.
The first is a register-derived conceptual graph connecting
municipalities through structural similarities rather than
geographical adjacency. Its attributes include administrative
classifications, altitude, surface area, population size,
urbanisation, and coastal location. These institutional descriptors
provide an operational representation of territorial affinity
without being selected through optimisation.
The second component is an external socio-economic field.
Social and economic indicators are combined into interpretable
composite dimensions, which are processed through Principal
Component Analysis (PCA). A weighted combination of the resulting
components defines the univariate field entering the energy
function alongside the interaction terms. 
Starting from the observed reference, local perturbations generate
configurations evaluated under each procedure's energy
specification. At a fixed comparison temperature, lower energy
corresponds to higher relative Gibbs density within that
specification. Proximity to the reference is assessed separately
through municipal deviations.
Simulated Annealing progressively reduces stochastic exploration
to investigate nearby alternatives rather than seek a global
minimum. The extent of the explored neighbourhood depends on the
transition rules, cooling schedule, and simulation duration.
Nearby lower-energy configurations provide the basis for studying
local variations, or ``vibrations'', and their association with
municipal socio-economic profiles, without assuming that the
observed state is an equilibrium or identifying a local minimum. 
Two complementary strategies implement this exploration.
The Continuous Ising procedure uses local Gaussian proposals
and a Metropolis-type acceptance rule. The Langevin-based procedure
combines gradient-driven increments with temperature-scaled noise,
followed by an acceptance step. Both use the same observed initial
configuration, graph, and external field, but differ in state
scaling and transition rules. Their comparison examines shared
and procedure-dependent empirical patterns without assuming
identical energy landscapes or a common stationary distribution.
Energy trajectories and municipal deviations characterise the
generated configurations. The retained ensembles and their
selection are specified in Section~\ref{sec4_3}.
Conformal calibration complements these comparisons through
municipality-specific intervals constructed by repeated
subsampling of the retained ensembles. Separate replicate subsets
are used for calibration and evaluation of coverage of the fixed
observed values. Interval widths and empirical coverage describe
territorial differences under this protocol; the procedure,
exchangeability argument, and scope of interpretation are
specified in Section~\ref{sec4_5}.
Together, these analyses provide an exploratory basis for relating
simulated territorial variation to register-derived socio-economic
profiles and formulating hypotheses about territorial organisation.


\section{Empirical application}\label{sec4}

This section illustrates the proposed methodology through an empirical
application focused on Italian municipalities. The analysis relies on the
\textit{ARCHIMEDE} statistical register maintained by the Italian
National Institute of Statistics (ISTAT), which integrates multiple
administrative sources into a comprehensive municipality-level database.
The empirical study investigates the territorial distribution of the
percentage of resident foreign population, adopting the workflow outlined in
Section~\ref{sec3} to explore the local energy landscape surrounding the
observed distribution.

\subsection{The input dataset}
\label{sec4_0}

The aforementioned complete register contains a total of $7908$
available municipalities with no missing values for the variables
considered in this study. 
The municipal indicators used in the empirical application
refer to 2021. 
For descriptive comparisons, the codes
of the official ISTAT territorial classification are aggregated
into two groups: codes A--D are labelled \emph{Central Hub},
and codes E--F are labelled \emph{Peripheral Area}.
These labels identify the aggregation adopted in this study.
The resulting grouping variable is kept exclusively for
descriptive purposes and is not involved in model construction.
The empirical application focuses on the North-East macro-region,
comprising $1385$ municipalities in the input dataset.
Two municipalities with no connections in the conceptual
interaction graph are excluded, leaving $1383$ municipalities
for the simulations.\footnote{The present
implementation represents the interaction graph through an
adjacency matrix for simplicity. The methodology is independent
of this storage representation. Sparse matrix or adjacency-list
implementations can reduce storage requirements when the graph
is sufficiently sparse, facilitating applications to larger
territorial systems, including the complete set of Italian
municipalities.}
This domain provides an empirical case study for illustrating
the energy-based framework.
The target variable is the percentage of resident foreign
population. To specify the socio-economic field acting on the
system, a set of base indicators was selected from the register
to build six composite indices (MPI1--MPI6), representing key
territorial dimensions. These indicators, along with their
respective polarities, are detailed in
Table~\ref{tab:mpi_detailed}.
\begin{table}[h]
\captionsetup{
  width=\textwidth,
  justification=justified,
  singlelinecheck=false
}
\caption{Description of Composite Indices}
\label{tab:mpi_detailed}
\centering
\begin{tabular}{p{0.1\linewidth}lp{0.25\linewidth}r}
\toprule
\textit{Composite Index}
& \textit{Base Indicator}
& \textit{Description}
& \textit{Polarity} \\
\midrule

\multirow{4}{*}{
  \begin{minipage}[t]{\linewidth}
  MPI1\\(Demographic Structure)
  \end{minipage}
}
& PERC\_ANZIANI
& \% of elderly ($65+$) in the population
& -1 \\
& PERC\_GIOVANI
& \% of young people ($<15$) in the population
& +1 \\
& PERC\_FAMIGLIE\_MINORI
& \% of households with minors
& +1 \\
& PERC\_FAM\_UNIPERSONALI\_ANZIANI
& \% of single-member elderly households
& -1 \\
\midrule

\multirow{3}{*}{
  \begin{minipage}[t]{\linewidth}
  MPI2\\(Cultural Level)
  \end{minipage}
}
& PERC\_NEET
& \% of NEETs (not in education, employment or training)
& -1 \\
& PERC\_LAUREATI
& \% of university graduates
& +1 \\
& PERC\_DIPLOMATI
& \% of high school graduates ($25$--$64$ years old)
& +1 \\
\midrule

\multirow{2}{*}{
  \begin{minipage}[t]{\linewidth}
  MPI3\\(Economic Well-being)
  \end{minipage}
}
& REDDITO\_MEDIANO\_EQUIVALENTE
& Median equivalent income (log-transformed)
& +1 \\
& PERC\_WORKINGPOOR
& \% of workers earning less than $60\%$ of the median wage
& -1 \\
\midrule

\multirow{3}{*}{
  \begin{minipage}[t]{\linewidth}
  MPI4\\(Employment Level)
  \end{minipage}
}
& PERC\_PRECARI
& \% of precarious workers (October snapshot)
& -1 \\
& PERC\_OCCUPATI
& \% of employed individuals ($20$--$64$ years old)
& +1 \\
& PERC\_FAM\_BASSA\_INTLAV
& \% of households with low work intensity
& -1 \\
\midrule

\multirow{3}{*}{
  \begin{minipage}[t]{\linewidth}
  MPI5\\(Territorial Attractiveness)
  \end{minipage}
}
& I\_ATTRAZIONE
& Attraction index
& +1 \\
& I\_AUTOCONTENIMENTO
& Self-containment index
& +1 \\
& I\_COESISTENZA
& Coexistence index
& +1 \\
\midrule

\multirow{4}{*}{
  \begin{minipage}[t]{\linewidth}
  MPI6\\(Population Dynamism)
  \end{minipage}
}
& STA
& Static individuals (no signs of work/study activity)
& -1 \\
& D\_INT
& Internal movers (within the same municipality)
& -1 \\
& D\_EST\_USCITA
& External movers leaving the municipality
& +1 \\
& D\_EST\_ENTRATA
& External movers entering the municipality
& +1 \\
\bottomrule
\end{tabular}
\end{table}
The assigned polarities orient the base indicators according to
the interpretation adopted for each composite dimension. Higher
composite values therefore express a stronger presence of the
characteristics assigned a positive polarity, after accounting
for negatively oriented indicators. These composite indices are
subsequently synthesised through a weighted combination of
principal components to construct the univariate external field
used in both stochastic procedures.
In parallel, territorial interactions are encoded using a
\emph{register-derived conceptual interaction graph}.
Municipalities sharing the same profile across the selected
institutional and structural attributes are connected with an
interaction weight equal to $1$. These links represent
similarity in the selected attributes rather than geographical
adjacency. The attributes used to establish the connections
are summarised in Table~\ref{tab_2}.
\begin{table}[h]
\captionsetup{
  width=\textwidth,
  justification=justified,
  singlelinecheck=false
}
\caption{Territorial attributes of municipalities}
\label{tab_2}
\centering
\begin{tabular}{p{0.25\linewidth} l p{0.6\linewidth}}
\toprule
\textit{Attribute}
& \textit{Item}
& \textit{Description} \\
\midrule

\multirow{3}{*}{
  \begin{minipage}[t]{\linewidth}
  ALT\\(Altitude of the centre)
  \end{minipage}
}
& 1 & Lowland (below 300 metres) \\
& 2 & Hill (between 300 and 600 metres) \\
& 3 & Mountain (above 600 metres) \\
\midrule

\multirow{3}{*}{
  \begin{minipage}[t]{\linewidth}
  POP\\(Resident Population Size)
  \end{minipage}
}
& 1 & Small (fewer than 5,000 inhabitants) \\
& 2 & Medium (between 5,000 and 50,000 inhabitants) \\
& 3 & Large (more than 50,000 inhabitants) \\
\midrule

\multirow{3}{*}{
  \begin{minipage}[t]{\linewidth}
  SUP\\(Surface Area)
  \end{minipage}
}
& 1 & Small (less than 15 km\textsuperscript{2}) \\
& 2 & Medium (between 15 and 100 km\textsuperscript{2}) \\
& 3 & Large (more than 100 km\textsuperscript{2}) \\
\midrule

\multirow{2}{*}{
  \begin{minipage}[t]{\linewidth}
  CLITO\\(Coastal Location)
  \end{minipage}
}
& 0 & Non-coastal \\
& 1 & Coastal \\
\midrule

\multirow{3}{*}{
  \begin{minipage}[t]{\linewidth}
  DEGURB\\(Urbanisation Level)
  \end{minipage}
}
& 1 & City / Densely populated areas \\
& 2 & Towns and suburbs / Intermediate density areas \\
& 3 & Rural areas / Sparsely populated areas \\
\bottomrule
\end{tabular}
\end{table}
This setup yields a simple undirected graph whose interaction
matrix $\mathbf{J}$ is a binary adjacency matrix with elements
in $\{0,1\}$. It specifies the pairwise interaction structure
introduced in Section~\ref{sec2_1} and provides a common
territorial representation for both stochastic procedures.
This conceptual representation aligns naturally with Official
Statistics, where administrative and structural classifications
provide established descriptors of territorial relationships
alongside geographical proximity.
For the Continuous Ising model, the observed target values are
linearly rescaled to $[-1,+1]$ for the energy calculations,
whereas Langevin dynamics operates directly on the original
$[0,100]$ percentage scale. Both procedures start from the same
observed municipal configuration, expressed on their respective
scales. Municipal results are reported on the percentage scale
to retain empirical interpretability. The common initial
reference and interaction graph do not imply an identical
numerical energy landscape for the two formulations.

\subsection{External field of the Ising model}
\label{sec4_2}

The exogenous socio-economic forcing acting on the territorial
system is represented by a scalar field value for each
municipality. The six composite indices ($MPI1$--$MPI6$)
describe interpretable dimensions of the municipal
socio-economic structure and exhibit non-negligible linear
dependencies. Principal Component Analysis (PCA) is used to
express this information through orthogonal components and
construct the external field introduced in Section~\ref{sec2}.
The same field is used in both stochastic procedures.
The degree of redundancy among the composite indices was
initially assessed using the Pearson correlation matrix
for the $1383$ municipalities included in the model.
\begin{table}[h]
\captionsetup{
  width=\textwidth,
  justification=justified,
  singlelinecheck=false
}
\caption{Correlation matrix of the composite indices}
\label{tab:corr_mpi}
\centering
\begin{tabular}{lcccccc}
\toprule
     & MPI1 & MPI2 & MPI3 & MPI4 & MPI5 & MPI6 \\
\midrule
MPI1 & 1.0000 & 0.1848 & 0.4334 & 0.5027 & 0.1294 & 0.2490 \\
MPI2 & 0.1848 & 1.0000 & 0.3758 & 0.1329 & 0.2434 & -0.1121 \\
MPI3 & 0.4334 & 0.3758 & 1.0000 & 0.6146 & 0.0993 & 0.1909 \\
MPI4 & 0.5027 & 0.1329 & 0.6146 & 1.0000 & -0.0194 & 0.2800 \\
MPI5 & 0.1294 & 0.2434 & 0.0993 & -0.0194 & 1.0000 & -0.1774 \\
MPI6 & 0.2490 & -0.1121 & 0.1909 & 0.2800 & -0.1774 & 1.0000 \\
\bottomrule
\end{tabular}
\end{table}
Table~\ref{tab:corr_mpi} shows positive associations among
several composite indices. The strongest correlation is
between $MPI3$ (economic well-being) and $MPI4$ (employment
level), with a coefficient of $0.6146$. Demographic structure
($MPI1$) is also positively associated with employment level
($0.5027$) and economic well-being ($0.4334$). Population
dynamism ($MPI6$) exhibits weaker associations with the other
dimensions, including a correlation of $0.2490$ with $MPI1$.
The six indices therefore represent related but distinct
dimensions of the municipal socio-economic profile. 
PCA is applied to the centred and standardised composite
indices. It transforms the correlated variables into
orthogonal components, providing a representation of their
joint variability. A weighted combination of the municipal
component scores defines the external field:
\begin{equation}\label{eq:ext_field}
  \textbf{h} = \lambda_{1}\textbf{pc}_{1}
  + \lambda_{2}\textbf{pc}_{2}
  + \ldots + \lambda_{6}\textbf{pc}_{6}
\end{equation}
where $\mathbf{h}=\{h_1,h_2,\ldots,h_n\}$ denotes the external
field associated with the $n$ municipalities composing the
territorial system. Each $\textbf{pc}_{k}$ contains the
municipal scores for the corresponding principal component.
The coefficients $\lambda_{k}$ are the normalised PCA
eigenvalues, equivalent to the proportions of total variance
explained by the components, and sum to one. These indexed
coefficients are distinct from the regularisation parameter
$\lambda$ in Equation~\ref{eq:hamilt}. A summary of the PCA
is reported in Table~\ref{tab:pca_summary}.
\begin{table}[h]
\captionsetup{
  width=\textwidth,
  justification=justified,
  singlelinecheck=false
}
\caption{Principal Component Analysis summary}
\label{tab:pca_summary}
\centering
\begin{tabular}{lcccccc}
\toprule
\textit{Component}
& \textit{PC1} & \textit{PC2} & \textit{PC3}
& \textit{PC4} & \textit{PC5} & \textit{PC6} \\
\midrule
Standard deviation
& 1.5134 & 1.1773 & 0.8914 & 0.8157 & 0.7342 & 0.5697 \\
Proportion of Variance
& 0.3817 & 0.2310 & 0.1324 & 0.1109 & 0.0898 & 0.0541 \\
Cumulative Proportion
& 0.3817 & 0.6127 & 0.7452 & 0.8561 & 0.9459 & 1.0000 \\
\bottomrule
\end{tabular}
\end{table}
The first three principal components account for $74.52\%$
of the total variance. The remaining components account for
$25.48\%$, with the sixth component contributing $5.41\%$.
All six components are retained in the construction of the
external field and contribute according to their respective
weights in Equation~\ref{eq:ext_field}. 
The resulting field provides a univariate synthesis of the
joint socio-economic profile used to evaluate configurations
under each energy specification. The complete PCA
representation preserves the variability of the input
indices, whereas their aggregation into a single field is a
modelling choice. Consequently, the field represents their
combined contribution to the stochastic exploration rather
than separate estimates of the effects of individual
socio-economic dimensions.

\subsection{MCMC simulations of the model}
\label{sec4_3}

The two stochastic procedures start from the observed municipal
configuration and use the same interaction graph and external
socio-economic field. Their energy specifications and state
transformations are detailed in Section~\ref{sec4_4}.
For each procedure, six independent chains are executed in
parallel on six workers. Each chain performs $100000$ attempted
single-municipality updates, yielding $600000$ attempts in total. 
At each iteration, one municipality is selected uniformly at
random. Both procedures use an initial temperature of $T_0=5000$
and the cooling rule $T=T_0/k$, where $k$ is initialised to $1$
and incremented only after an accepted proposal. The temperature
therefore remains unchanged following a rejection, and each
chain follows its own acceptance-dependent cooling schedule. 
For the Continuous Ising procedure, the selected municipal
percentage receives a Gaussian increment with mean zero and
standard deviation $0.01$ percentage points. The proposed value
is clipped to $[0,100]$ before its transformed value is used
to evaluate the energy difference. 
For the Langevin-based procedure, the gradient on the percentage
scale is $\mathbf{g}=J\mathbf{s}+\mathbf{h}$. At each iteration,
the adaptive step is $a=2T/v_g$, where $v_g$ is the sample
variance of the gradient components across municipalities.
The proposed increment for the selected municipality $i$ is
$-a g_i+\sqrt{2Ta}\,\eta$, with
$\eta\sim\mathcal{N}(0,1)$. The resulting municipal value is
also clipped to $[0,100]$. This implementation updates one
coordinate at a time and combines a state-dependent step
with a subsequent energy-based acceptance rule. 
In both procedures, acceptance uses the probability
$\min\{1,\exp(-\Delta H/T)\}$, where $\Delta H$ is evaluated
for the proposal after clipping under the corresponding
energy specification. In particular, the Langevin energy
increment uses the effective change in the municipal value
after clipping. A rejected proposal leaves the configuration
unchanged. The current configuration is stored after every
attempt, so repeated configurations following rejections
remain in the simulation output. 
The implemented acceptance rule does not include a
proposal-density correction for the gradient-driven
Langevin proposal or for clipping at the boundaries.
The procedures are consequently used as stochastic
exploration mechanisms; exact sampling from a Gibbs
distribution is not asserted. The fixed-temperature Gibbs
density ratios introduced in Section~\ref{sec4_4} provide
a separate interpretation of the recorded energy differences. 
The outputs are concatenated in chain order for storage and
subsequent selection. This concatenation does not constitute
a single continuous Markov trajectory.
The last $75000$ configurations of the concatenated output
are retained for the descriptive and energy comparisons,
while the last $50000$ form the subsampling pool for the
conformal analysis. Under this selection rule, both ensembles
are drawn entirely from the sixth chain. The reported ensemble
summaries therefore describe the retained segment of that
chain rather than a pooled sample from all six chains.

\subsection{Energy and likelihood of the simulated configurations}
\label{sec4_4}

The behaviour of both stochastic dynamics was investigated
through the energy and log-likelihood of the configurations
sampled from the retained final segment of their simulation
outputs. These quantities describe the energetic position of
the simulated territorial configurations relative to the
observed reference under each formulation. They complement
the assessment of municipal deviations from the observed
values in characterising the explored neighbourhood. 
The empirical implementation uses the following energy
specifications:
\begin{equation}\label{eq:implemented_energies}
\begin{aligned}
H_{\mathrm{CI}}(\mathbf{s})
  &= \tfrac{1}{2}\mathbf{z}^{\mathsf{T}}\mathbf{J}\mathbf{z}
     + \mathbf{h}^{\mathsf{T}}\mathbf{z}, \\
H_{\mathrm{LD}}(\mathbf{s})
  &= \tfrac{1}{2}\mathbf{s}^{\mathsf{T}}\mathbf{J}\mathbf{s}
     + \mathbf{h}^{\mathsf{T}}\mathbf{s}.
\end{aligned}
\end{equation}
where $\mathbf{s}$ contains municipal percentages and
$\mathbf{z}$ is their linear transformation, defined by
\begin{equation}\label{eq:ising_state_scaling}
z_i =
2\,\frac{s_i-\min_j s_j^{\mathrm{ref}}}
        {\max_j s_j^{\mathrm{ref}}-\min_j s_j^{\mathrm{ref}}}
-1.
\end{equation}
The reference minimum and maximum are fixed throughout
the simulation. This transformation maps the observed
configuration to $[-1,+1]$; simulated values on the transformed
scale are not restricted to this interval. Municipal states
are constrained to $[0,100]$ in both procedures. 
These empirical specifications use positive signs for the
interaction and external-field terms and omit the quadratic
regularisation term appearing in the theoretical formulation
of Equation~\ref{eq:hamilt}. The same interaction matrix
$\mathbf{J}$ and external field $\mathbf{h}$ enter both
specifications, but their different state scales produce
different energy functions on the municipal percentage scale. 
For each sampled configuration $\mathbf{s}$, the ratio between
its Hamiltonian $H(\mathbf{s})$ and the Hamiltonian of the
reference configuration $H_{ref}$ was computed using the
corresponding energy specification. Since the reference
energies are positive in this application, energy ratios
smaller than $1$ indicate configurations that are energetically
more favourable than the observed one under that specification.
Their lower energy does not, by itself, establish convergence
to a local minimum or proximity to the observed configuration. 
The relative Gibbs density of each simulated configuration
was also compared with that of the reference configuration.
According to Equation~\ref{eq:boltzmann}, at a common fixed
temperature within each formulation, the log-likelihood ratio
is given by:
\begin{equation}\label{eq:ll_ratio}
\log \Big(\frac{P(\mathbf{s})}{P(\mathbf{s}_{ref})}\Big)
\propto -\Delta H
\end{equation}
where $\Delta H=H(\mathbf{s})-H_{ref}$ denotes the energy
difference between the simulated and the reference
configuration. 
The partition function cancels in the ratio,
avoiding its explicit evaluation.\footnote{At a fixed
temperature $T>0$, the log-likelihood ratio equals
$-\Delta H/T$. The temperature used for this comparison
is held constant across configurations and is distinct
from the changing temperature of the annealing schedule.}
The reported comparisons use $T=1$ for both formulations. 
For notational convenience, the logarithm of the relative Gibbs
density defined in Equation~\ref{eq:ll_ratio} will be denoted by
$\mathcal{LL}_{\mathrm{ratio}}$ throughout the remainder of the
paper, consistently with Section~\ref{sec2_1}. Positive values
indicate a higher Gibbs density than that of the observed
reference under the corresponding energy specification.
Because this quantity is determined by the energy difference
at the chosen temperature, it provides a probabilistic
interpretation of the energetic ordering rather than an
independent assessment of model fit. 
These relative quantities $H/H_{\mathrm{ref}}$ and the
corresponding log-likelihood ratio are considered in the
subsequent analysis. Comparisons with the reference are
performed separately for each formulation; the absolute
magnitudes of the log-likelihood ratios are not interpreted
as evidence that one stochastic strategy is preferable
to the other.

\subsection{Uncertainty quantification of the model}
\label{sec4_5}

The uncertainty analysis uses repeated random subsampling of the
retained final ensemble from each stochastic procedure. Let $n$
denote the number of municipalities. For each replicate
$b=1,\ldots,B$, a subset of $m$ configurations is drawn without
replacement from the retained ensemble. Subsets are drawn
independently across replicates and may therefore overlap.
For each municipality $i$, the empirical quantiles at levels
$\alpha/2$ and $1-\alpha/2$ are computed from its $m$ sampled
values, yielding initial bounds $L_{bi}$ and $U_{bi}$.
Repeating this procedure produces two $B \times n$ matrices
of interval bounds. These bounds are obtained directly from
the simulated configurations, without fitting a quantile
regression model. 
Replicates are partitioned into disjoint calibration and
validation subsets, comprising $20\%$ and $80\%$ of the
replicates, respectively. The calibration target is the
observed municipal value $s_i^{\mathrm{ref}}$, which remains
fixed across replicates. For each calibration replicate,
the nonconformity score is
\begin{equation}
r_{bi} =
\max\left\{
L_{bi}-s_i^{\mathrm{ref}},
s_i^{\mathrm{ref}}-U_{bi}
\right\}.
\label{eq:subsampling_score}
\end{equation}
Writing $B_{\mathrm{cal}}$ for the number of calibration
replicates, the municipality-specific adjustment
$\widehat q_i$ is computed as the empirical quantile of these
scores at level
$\lceil(B_{\mathrm{cal}}+1)(1-\alpha)\rceil/B_{\mathrm{cal}}$.
Both the initial bounds and the calibration adjustment are
computed using the linear interpolation implemented by
\texttt{quantile(..., type = 7)} in R. The adjustment therefore
uses the finite-sample quantile level with interpolation,
rather than the direct order-statistic selection described
for standard split CQR in Section~\ref{sec2_5}.
If the specified level exceeds one, the adjustment is set
to positive infinity. 
Scores may be negative when the reference value lies inside
the initial interval. Consequently, calibration may either
expand or contract the initial bounds. For each validation
replicate, the adjusted interval is
\begin{equation}
C_{bi} =
\left[
L_{bi}-\widehat q_i,\,
U_{bi}+\widehat q_i
\right]\cap[0,100],
\label{eq:subsampling_interval}
\end{equation}
where the restriction reflects the percentage scale of the
target variable. If the adjusted lower bound exceeds the
adjusted upper bound, the interval is empty. Empty intervals
are retained in the validation summaries and contribute
zero width and zero coverage. 
Conditional on the retained ensemble and the observed
reference configuration, independently generated subsets
follow the same sampling rule. The resulting interval bounds
and nonconformity scores are therefore exchangeable across
replicates for each municipality. This argument concerns the
random subsampling procedure; it does not require successive
states of the original simulation to be independent, nor does
it assume independence between municipalities. 
For each municipality, empirical coverage is the proportion
of validation intervals containing its fixed observed value:
\begin{equation}
\widehat{\mathrm{coverage}}_i =
\frac{1}{B_{\mathrm{val}}}
\sum_{b\in\mathcal{V}}
\mathbf{1}
\left\{
s_i^{\mathrm{ref}}\in C_{bi}
\right\},
\label{eq:subsampling_coverage}
\end{equation}
where $\mathcal{V}$ denotes the validation subset and
$B_{\mathrm{val}}$ its size. Adaptivity is measured by the
mean width of the adjusted intervals across all validation
replicates, including those producing empty intervals. 
The procedure thus assesses how consistently intervals
constructed from randomly selected simulated configurations,
after calibration, contain the observed municipal reference.
Its interpretation is conditional on the retained ensemble
and the specified subsampling protocol. Coverage concerns
the fixed reference under repeated subsampling, rather than
future observations or simultaneous coverage of the entire
territorial configuration. Because the observed reference
also enters calibration, this diagnostic does not constitute
an independent validation of the territorial model. The
interval widths quantify the extent of the calibrated ranges
and provide a descriptive basis for comparing municipalities
under the same simulation and calibration protocol. 
Thick borders in the uncertainty maps identify local
concentrations of larger calibrated interval widths.
These are detected using standardised Getis--Ord $G_i$
scores above $1.96$, with row-standardised weights based
on geographical queen contiguity. The highlighted areas
provide an exploratory spatial diagnostic; thresholds
are not adjusted for multiple comparisons.


\section{Results}

\begin{figure}[h]
  \centering
  \captionsetup{
    width=\textwidth,
    justification=justified,
    singlelinecheck=false
  }
  \caption{Recorded energy trajectories of the Continuous
  Ising model. Vertical dotted lines mark chain boundaries;
  the concatenated configuration index is shown on a
  logarithmic scale.}
  \label{fig_1}
  \includegraphics[width=\textwidth]{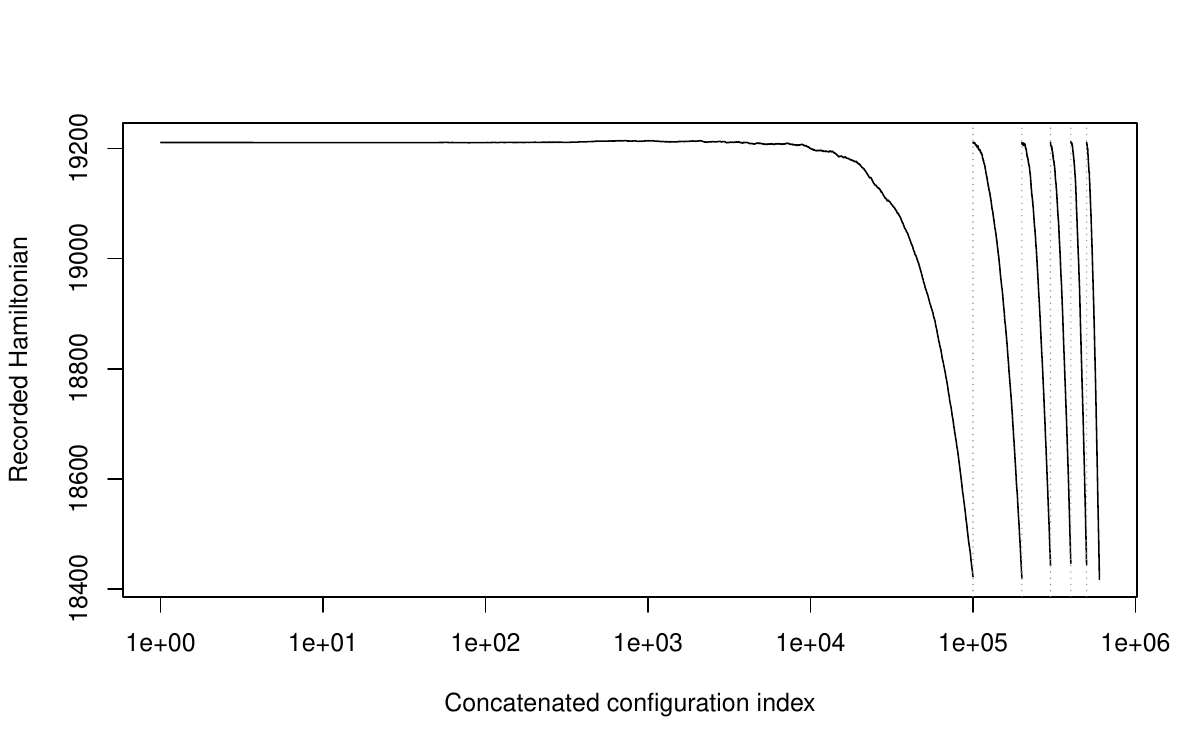}
\end{figure}

\begin{figure}[h]
  \centering
  \captionsetup{
    width=\textwidth,
    justification=justified,
    singlelinecheck=false
  }
  \caption{Recorded energy trajectories of the Langevin-based
  procedure. Vertical dotted lines mark chain boundaries;
  the concatenated configuration index is shown on a
  logarithmic scale.}
  \label{fig_2}
  \includegraphics[width=\textwidth]{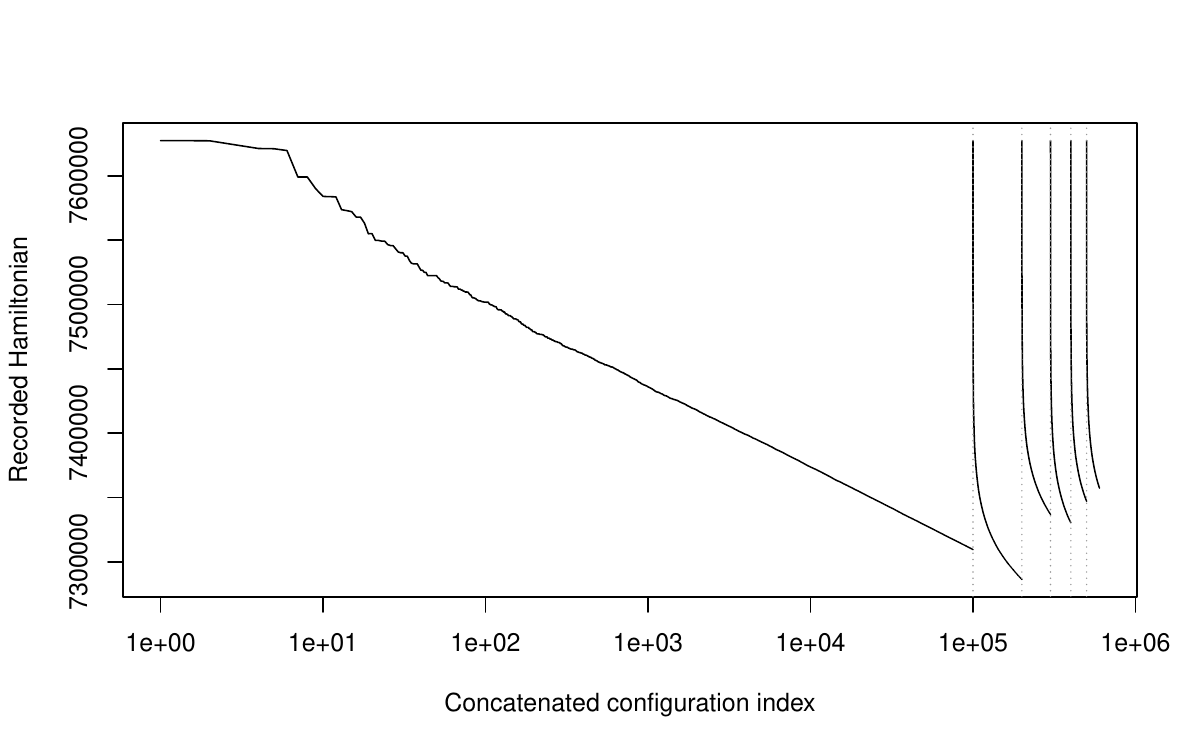}
\end{figure}

The results concern the ensembles selected according to
Section~\ref{sec4_3}. Simulations used six workers on a laptop
equipped with an 11th generation Intel Core i5-1145G7 processor
($4$ physical cores, $8$ logical threads, $2.60$ GHz) and $16$ GB of RAM,
running 64-bit Microsoft Windows. The simulation stage required
approximately $1.91$ minutes for the Continuous Ising model
and $14.44$ minutes for the Langevin-based procedure,
excluding energy audits, conformal calibration, and production
of tables and figures. Simulation settings were specified
without a formal sensitivity analysis. 
Figures~\ref{fig_1} and~\ref{fig_2} show the recorded energy
trajectories. The initial reference energies are
$H_{\mathrm{ref}}=19211.12$ for the Continuous Ising model
and $H_{\mathrm{ref}}=7627322.41$ for the Langevin-based
procedure, reflecting the respective state scales and
energy specifications defined in Section~\ref{sec4_4}.
The Continuous Ising trajectory initially remains close
to the reference energy, with small fluctuations, before
declining more markedly. Terminal energies across the six
chains range approximately from $18400$ to $18500$.
For the Langevin-based procedure, the decline appears earlier
in the first chain, with terminal energies between
approximately $7.26 \times 10^6$ and $7.36 \times 10^6$.
Both procedures therefore reach energies below their
respective observed references. 
Returns to the reference energy at the marked boundaries
identify the starts of independent chains. The logarithmic
horizontal axis expands the early part of the concatenated
output and compresses later chains despite their equal
numbers of updates. Decreasing energy alone does not
establish stationarity or convergence to a local minimum;
proximity to the observed configuration is assessed through
the municipal comparisons below. For every stored
configuration in both procedures, recorded energies agreed
with direct Hamiltonian evaluations within the numerical
tolerance used in the audit.

\subsection{Observed and locally explored configurations}

The coherence between the observed territorial configuration
($\mathbf{s}_{\mathrm{ref}}$) and the average configurations
generated by the Continuous Ising model
($\mathbf{s}^{CI}_{\mathrm{stat}}$) and the Langevin dynamics
($\mathbf{s}^{LD}_{\mathrm{stat}}$) was assessed using the last
$75000$ configurations of the concatenated output for each
procedure. For each municipality, the corresponding component
of the average configuration is the mean value of the target
variable over the retained configurations. The subscript
$\mathrm{stat}$ identifies these simulation summaries and
does not imply that stationarity has been established.
Municipal simulation averages are rounded to two decimal
places before computing the comparisons reported in
Table~\ref{tab:ising_langevin_comparison}.
\begin{table}[ht]
\centering
\captionsetup{
  width=\textwidth,
  justification=justified,
  singlelinecheck=false
}
\caption{Observed and simulated configurations:
descriptive comparison}
\label{tab:ising_langevin_comparison}
\begin{tabular}{lcc}
\toprule
\textit{Statistic / Test}
& \textit{Continuous Ising model}
& \textit{Langevin dynamics} \\
\midrule
\multicolumn{3}{l}{\textit{Descriptive statistics}} \\
Initial Mean   & 7.9339 & 7.9339 \\
Estimated Mean & 7.9805 & 7.8360 \\
\midrule
\multicolumn{3}{l}{\textit{Shapiro--Wilk normality tests}} \\
Observed ($W$, p-value)
& 0.9715, $6.90\times 10^{-16}$
& 0.9715, $6.90\times 10^{-16}$ \\
Estimated ($W$, p-value)
& 0.9715, $6.68\times 10^{-16}$
& 0.9729, $1.89\times 10^{-15}$ \\
\midrule
\multicolumn{3}{l}{\textit{Paired $t$-test}} \\
$t$ (df)
& $-27.94$ (1382)
& $14.01$ (1382) \\
p-value
& $< 2.2\times 10^{-16}$
& $< 2.2\times 10^{-16}$ \\
95\% CI
& $[-0.0499, -0.0434]$
& $[0.0842, 0.1116]$ \\
Mean diff.
& $-0.0466$
& $0.0979$ \\
\midrule
\multicolumn{3}{l}{\textit{Paired Wilcoxon test}} \\
$V$, p-value
& 124051, $< 2.2\times 10^{-16}$
& 841414, $< 2.2\times 10^{-16}$ \\
\midrule
\multicolumn{3}{l}{\textit{Distance measures}} \\
MAE         & 0.0639 & 0.1016 \\
RMSE        & 0.0776 & 0.2775 \\
Correlation & 0.9999 & 0.9979 \\
\bottomrule
\end{tabular}
\end{table}
The descriptive statistics reported in
Table~\ref{tab:ising_langevin_comparison} indicate that the
average configurations obtained from both stochastic
procedures remain close to the observed territorial
configuration. The Continuous Ising model produces an
overall mean of $7.9805\%$, slightly above the observed
mean of $7.9339\%$, whereas Langevin dynamics yields
a lower average of $7.8360\%$.
The mean differences in the table are calculated as
observed minus simulated values. The term ``Estimated''
identifies summaries of the simulated municipal averages. 
The Shapiro--Wilk tests return small nominal p-values for
both the observed municipal values and the simulated
municipal averages, indicating departures from normality
under the conventional test assumptions. The paired
$t$-test and Wilcoxon test likewise return small nominal
p-values, while the mean differences remain limited in
magnitude: $-0.0466$ percentage points for the Continuous
Ising model and $0.0979$ for Langevin dynamics.
Because municipalities are linked through the interaction
graph, these conventional tests do not account for the
dependence represented in the model. Their p-values and
the reported $t$-test confidence intervals are therefore
retained as supplementary comparisons rather than used
as dependence-adjusted inferential evidence. 
The reported MAE and RMSE quantify the distance between
the observed municipal values and the corresponding
simulation averages, expressed in percentage points.
Both measures are smaller for the Continuous Ising model:
the MAE is $0.0639$, compared with $0.1016$ for Langevin
dynamics, and the RMSE is $0.0776$, compared with $0.2775$.
The larger separation between MAE and RMSE for Langevin
dynamics indicates a greater influence of large municipal
deviations on its squared-error summary.
The correlations with the observed municipal values are
$0.9999$ and $0.9979$, respectively. 
These quantities are descriptive measures of agreement
with the initial reference rather than measures of
predictive performance. They concern ensemble averages
and do not establish the proximity of every individual
simulated configuration. The comparison of municipal
averages and the subsequent conformal analysis therefore
address distinct features of the retained simulation output.

\subsection{Energy and likelihood of the sampled configurations}

The recorded energy of the observed reference configuration is
$H_{\mathrm{ref}}=19211.12$ for the Continuous Ising model and
$H_{\mathrm{ref}}=7627322.41$ for Langevin dynamics.
These values reflect the different state scales used in the
two implementations: normalised states for the Continuous
Ising energy calculation and percentage values for Langevin
dynamics, according to Section~\ref{sec4_4}. 
For each procedure, the analysis considers the last $75000$
configurations of the concatenated simulation output.
The energy of each retained configuration is expressed
relative to its corresponding reference value through
$H/H_{\mathrm{ref}}$. The log-likelihood ratio is evaluated
according to Equation~\ref{eq:ll_ratio}, with the comparison
temperature fixed at $T=1$.
Table~\ref{tab_4} summarises the resulting distributions.
\begin{table}[h]
\centering
\captionsetup{
  width=\textwidth,
  justification=justified,
  singlelinecheck=false
}
\caption{Energy and log-likelihood ratio}
\label{tab_4}
\begin{tabular}{l cc cc}
\toprule
\multirow{2}{*}{\textit{Statistic}}
  & \multicolumn{2}{c}{\textit{Continuous Ising model}}
  & \multicolumn{2}{c}{\textit{Langevin dynamics}} \\
\cmidrule(lr){2-3} \cmidrule(lr){4-5}
  & $H/H_{\text{ref}}$ & $\mathcal{LL}_{\mathrm{ratio}}$
  & $H/H_{\text{ref}}$ & $\mathcal{LL}_{\mathrm{ratio}}$ \\
\midrule
Min.    & $0.9587$ & $87.15$  & $0.9646$ & $231055$ \\
1st Qu. & $0.9696$ & $220.47$ & $0.9654$ & $246818$ \\
Median  & $0.9800$ & $383.43$ & $0.9663$ & $256803$ \\
Mean    & $0.9789$ & $404.94$ & $0.9666$ & $254871$ \\
3rd Qu. & $0.9885$ & $583.95$ & $0.9676$ & $264141$ \\
Max.    & $0.9955$ & $793.81$ & $0.9697$ & $269904$ \\
\bottomrule
\end{tabular}
\end{table}
All retained configurations included in Table~\ref{tab_4}
have recorded energies below their respective reference
values ($H/H_{\mathrm{ref}}<1$).
The selected simulation segments therefore contain
configurations that are energetically more favourable than
the observed reference under the corresponding energy
evaluation. This finding does not, by itself, determine
whether the observed configuration is a local minimum. 
The Continuous Ising energy ratios have a median of $0.9800$
and an interquartile range of $0.0189$.
The corresponding Langevin values are $0.9663$ and $0.0023$.
Relative to their respective reference energies, the retained
Langevin configurations thus exhibit a larger median
proportional energy reduction and a smaller dispersion
of energy ratios.
These summaries describe variation in recorded energy;
they do not directly measure distance from the observed
territorial configuration, the width of an energy basin,
or the speed of convergence. 
The log-likelihood ratios are positive throughout both
retained ensembles. At the fixed comparison temperature,
this corresponds to higher relative Gibbs density than
at the observed reference, according to
Equation~\ref{eq:ll_ratio}.
The log-likelihood ratios express the same energy differences
on an additive scale and therefore complement, rather than
independently validate, the energy-ratio summaries.
Each column of Table~\ref{tab_4} summarises its respective
distribution separately; since the log-likelihood ratio
decreases as energy increases, its minimum corresponds
to the maximum energy ratio, and conversely. 
The different numerical scales preclude direct comparison
of the absolute energy and log-likelihood values across
the two implementations. Energy ratios provide a dimensionless
description relative to each reference, but do not establish
equivalence of the underlying energy evaluations.
The results document distinct distributions of relative
energy in the retained simulation segments, to be considered
alongside the municipal comparisons and conformal
uncertainty analysis.

\subsection{Conformal uncertainty analysis}
\label{sec:CP_res}

\begin{figure}[h]
  \centering
  \captionsetup{
    width=\textwidth,
    justification=justified,
    singlelinecheck=false
  }
  \caption{Conformal uncertainty map of the Continuous
  Ising model}
  \label{fig:umap_1}
  \includegraphics[width=\textwidth]
    {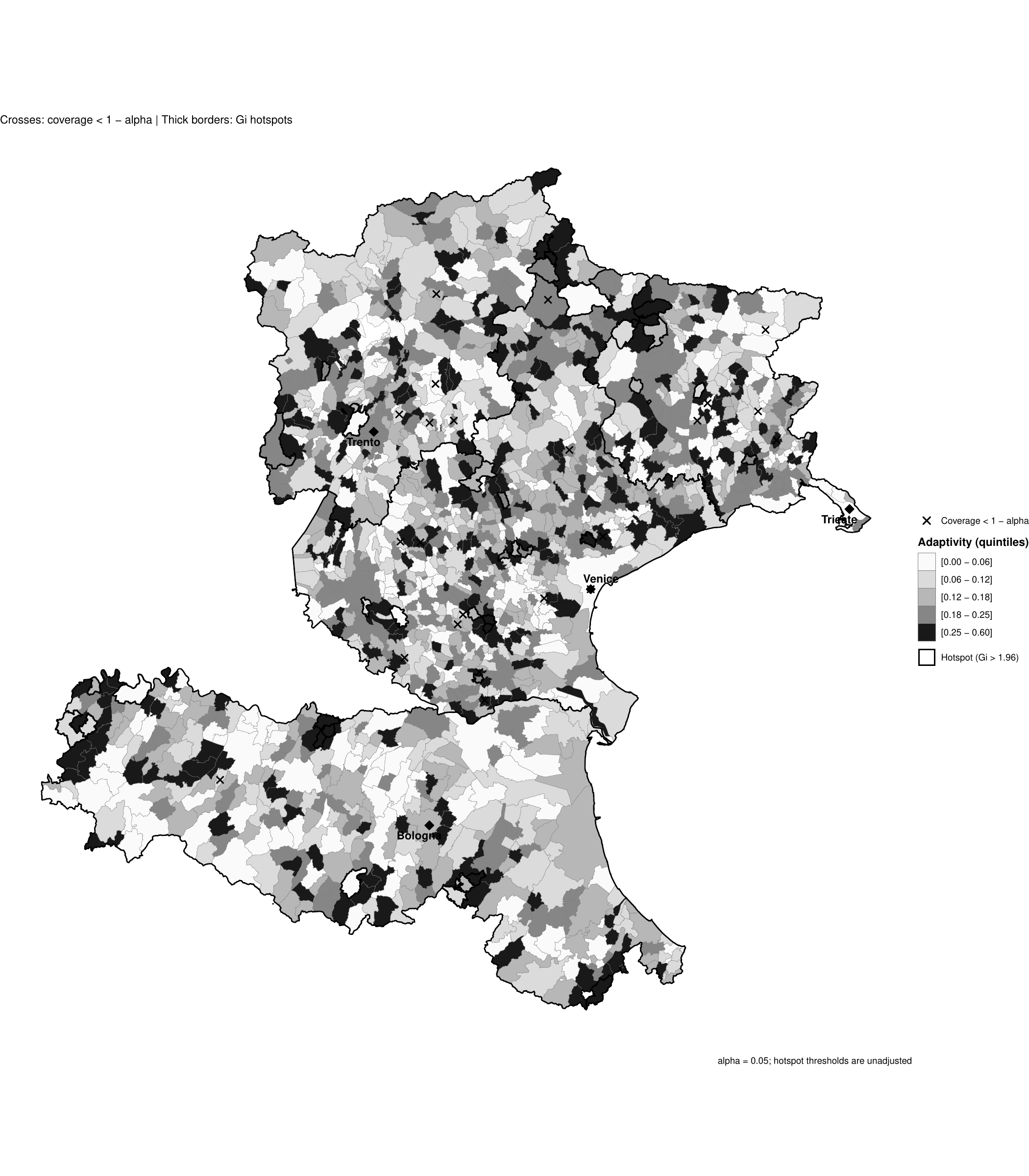}
\end{figure}

\begin{figure}[h]
  \centering
  \captionsetup{
    width=\textwidth,
    justification=justified,
    singlelinecheck=false
  }
  \caption{Conformal uncertainty map of the Langevin dynamics}
  \label{fig:umap_2}
  \includegraphics[width=\textwidth]
    {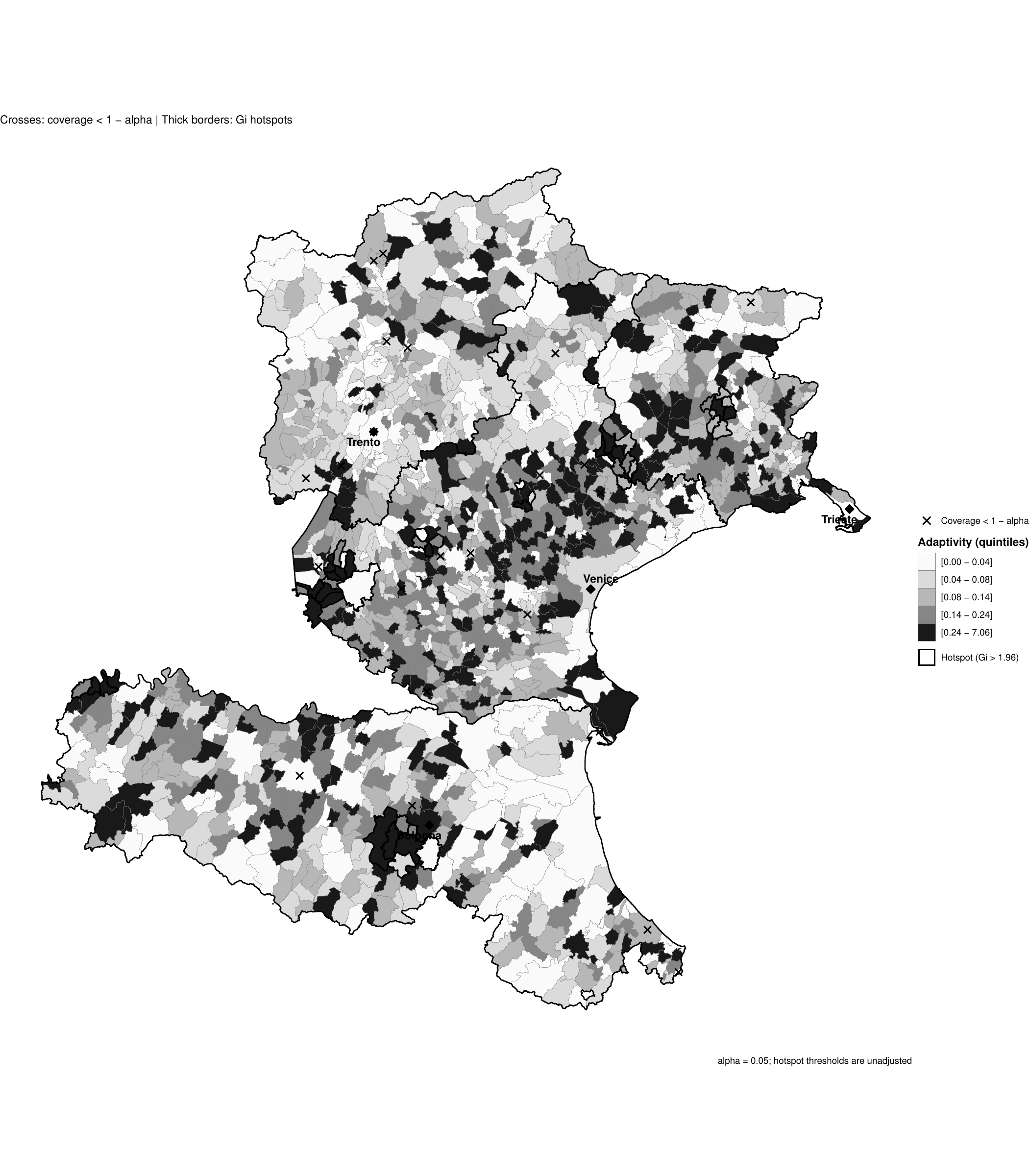}
\end{figure}

The procedure described in Section~\ref{sec4_5} was applied
with $\alpha=0.05$, using the last $N=50000$ configurations
to generate $B=10000$ independent subsampling replicates,
each containing $m=200$ configurations selected without
replacement. The last $2000$ replicates were used for
calibration and the first $8000$ for validation.
\begin{table}[!t]
\centering
\captionsetup{
  width=\textwidth,
  justification=justified,
  singlelinecheck=false
}
\caption{Coverage and adaptivity of the conformal uncertainty
analysis. Adaptivity is expressed in percentage points.}
\label{tab:coverage_adaptivity}
\begin{tabular}{lrrrrrr}
\toprule
 & Min. & 1st Qu. & Median & Mean & 3rd Qu. & Max. \\
\midrule
\multicolumn{7}{l}{\textit{Coverage}} \\
Continuous Ising
 & 0.9398 & 0.9712 & 0.9938 & 0.9852 & 0.9999 & 1.0000 \\
Langevin dynamics
 & 0.9426 & 0.9700 & 0.9925 & 0.9843 & 0.9996 & 1.0000 \\
\midrule
\multicolumn{7}{l}{\textit{Adaptivity}} \\
Continuous Ising
 & 0.0001 & 0.0754 & 0.1483 & 0.1611 & 0.2315 & 0.5987 \\
Langevin dynamics
 & 0.0000 & 0.0463 & 0.1059 & 0.1998 & 0.2122 & 7.0571 \\
\bottomrule
\end{tabular}
\end{table}
Table~\ref{tab:coverage_adaptivity} summarises municipal
coverage and adaptivity. Average coverage is $0.9852$
for Continuous Ising and $0.9843$ for Langevin dynamics,
above the nominal level of $0.95$. Nevertheless, each
procedure has $23$ of the $1383$ municipalities below
this level, with minimum coverage of $0.9398$ and
$0.9426$, respectively. 
Mean adaptivity is $0.1611$ percentage points for
Continuous Ising and $0.1998$ for Langevin dynamics,
whereas the medians are $0.1483$ and $0.1059$.
The respective maxima are $0.5987$ and $7.0571$.
Langevin dynamics thus combines a lower median with
a more pronounced upper tail and a higher mean;
neither procedure produces uniformly narrower intervals. 
The minimum Langevin adaptivity, displayed as $0.0000$,
is positive before rounding, approximately
$2.83 \times 10^{-5}$ percentage points. Calibration
produced $2083$ empty validation intervals for Continuous
Ising and $3$ for Langevin dynamics, out of $11064000$
per procedure. These contribute zero width and zero
coverage, as specified in Section~\ref{sec4_5}.
The diagnostics concern the combined simulation and
calibration procedure: widths do not directly measure
energy-basin geometry, and coverage concerns the fixed
observed reference under repeated subsampling. 
Figures~\ref{fig:umap_1} and~\ref{fig:umap_2} display
municipal adaptivity and identify municipalities with
empirical coverage below $1-\alpha$.
Each map uses its own quintile classes: matching colours
indicate relative positions within each distribution,
not equal interval widths. Comparisons between procedures
therefore require the numerical legend values.


\section{Analysis of the socio-economic aspects}

The socio-economic analysis relates simulated municipal
deviations and calibrated interval widths to territorial
attributes and composite-index profiles. These comparisons
describe associations under the combined interaction
structure and external field; they do not isolate the
contribution of individual indices. 
The results of both models are summarised in
Table~\ref{tab:c1_summary_all},
Table~\ref{tab:c1_mpi_all},
Table~\ref{tab:c2_summary_all} and
Table~\ref{tab:c2_mpi_all}. Municipalities are grouped by
empirical coverage status, municipality type (\emph{type}),
and territorial attribute class (\emph{class}).
The two municipality types correspond to the aggregation used
in the analysis: codes A--D form the first group
(\emph{Centr. Hub}), and codes E--F form the second
(\emph{Periph. Area}). Territorial attribute classes follow
Table~\ref{tab_2}.
Municipalities with empirical coverage at least $1-\alpha$
are classified as \emph{Covered}; those below this threshold
are classified as \emph{Not covered}. Classification uses
unrounded municipal coverage values. 
The reported statistics include the number of municipalities
in each group (\emph{n}), the mean values of the uncertainty
diagnostics (\emph{coverage} and \emph{adaptivity}), the observed
and simulated mean percentages of resident foreign population
($y_{ref}$ and $y_{est}$), their relative difference
$\delta = 100 \cdot (y_{est}/y_{ref} - 1)$, and the mean values
of the six composite socio-economic indices (MPI1--MPI6).
Adaptivity is expressed in percentage points, whereas
$\delta$ is a percentage change relative to the observed
group mean. 
The composite indices are fixed municipal characteristics.
The same municipalities are classified separately under each
territorial attribute, so counts must not be summed across
attribute blocks. Only non-empty groups are reported.
Differences between their group averages across the two
procedures reflect differences in the municipalities assigned
to each coverage category, rather than changes in the indices
generated by the simulations.
\begin{table}[ht]
\centering
\tiny
\captionsetup{
  width=\textwidth,
  justification=justified,
  singlelinecheck=false
}
\caption{Continuous Ising: summary results by territorial attributes}
\label{tab:c1_summary_all}
\begin{tabular}{l r r r r r r r}
\toprule
\textit{type} & \textit{class} & \textit{n}
& \textit{coverage} & \textit{adaptivity}
& $y_{ref}$ & $y_{est}$ & $\delta$ (\%) \\
\midrule
\multicolumn{8}{l}{\textit{Results by ALT}}\\
\multicolumn{8}{l}{\textit{Covered}}\\
Centr. Hub & 1 & 887 & 0.986 & 0.160
 & 9.029 & 9.077 & 0.532 \\
Centr. Hub & 2 & 139 & 0.985 & 0.127
 & 6.418 & 6.440 & 0.339 \\
Centr. Hub & 3 & 110 & 0.987 & 0.184
 & 5.767 & 5.819 & 0.889 \\
Periph. Area & 1 & 9 & 0.988 & 0.150
 & 8.928 & 8.943 & 0.174 \\
Periph. Area & 2 & 29 & 0.987 & 0.147
 & 6.341 & 6.373 & 0.495 \\
Periph. Area & 3 & 186 & 0.986 & 0.183
 & 5.422 & 5.482 & 1.093 \\
\multicolumn{8}{l}{\textit{Not covered}}\\
Centr. Hub & 1 & 15 & 0.947 & 0.181
 & 7.949 & 8.012 & 0.797 \\
Centr. Hub & 2 & 1 & 0.949 & 0.055
 & 3.230 & 3.240 & 0.310 \\
Centr. Hub & 3 & 5 & 0.945 & 0.096
 & 5.992 & 6.000 & 0.134 \\
Periph. Area & 3 & 2 & 0.948 & 0.235
 & 6.020 & 6.115 & 1.578 \\
\midrule
\multicolumn{8}{l}{\textit{Results by POP}}\\
\multicolumn{8}{l}{\textit{Covered}}\\
Centr. Hub & 1 & 592 & 0.985 & 0.161
 & 7.094 & 7.139 & 0.643 \\
Centr. Hub & 2 & 523 & 0.986 & 0.158
 & 9.644 & 9.691 & 0.479 \\
Centr. Hub & 3 & 21 & 0.985 & 0.094
 & 13.898 & 13.903 & 0.038 \\
Periph. Area & 1 & 209 & 0.986 & 0.184
 & 5.510 & 5.568 & 1.045 \\
Periph. Area & 2 & 15 & 0.980 & 0.072
 & 8.079 & 8.081 & 0.025 \\
\multicolumn{8}{l}{\textit{Not covered}}\\
Centr. Hub & 1 & 11 & 0.947 & 0.137
 & 5.774 & 5.814 & 0.693 \\
Centr. Hub & 2 & 9 & 0.946 & 0.189
 & 8.217 & 8.278 & 0.744 \\
Centr. Hub & 3 & 1 & 0.949 & 0.037
 & 14.960 & 14.970 & 0.067 \\
Periph. Area & 1 & 1 & 0.950 & 0.217
 & 7.420 & 7.500 & 1.078 \\
Periph. Area & 2 & 1 & 0.947 & 0.252
 & 4.620 & 4.730 & 2.381 \\
\midrule
\multicolumn{8}{l}{\textit{Results by SUP}}\\
\multicolumn{8}{l}{\textit{Covered}}\\
Centr. Hub & 1 & 216 & 0.986 & 0.129
 & 7.352 & 7.372 & 0.279 \\
Centr. Hub & 2 & 837 & 0.986 & 0.172
 & 8.590 & 8.645 & 0.645 \\
Centr. Hub & 3 & 83 & 0.984 & 0.090
 & 9.125 & 9.131 & 0.061 \\
Periph. Area & 1 & 27 & 0.989 & 0.103
 & 5.614 & 5.634 & 0.350 \\
Periph. Area & 2 & 159 & 0.986 & 0.204
 & 5.782 & 5.854 & 1.245 \\
Periph. Area & 3 & 38 & 0.982 & 0.113
 & 5.312 & 5.315 & 0.045 \\
\multicolumn{8}{l}{\textit{Not covered}}\\
Centr. Hub & 1 & 2 & 0.945 & 0.091
 & 5.710 & 5.715 & 0.088 \\
Centr. Hub & 2 & 17 & 0.947 & 0.175
 & 7.224 & 7.281 & 0.790 \\
Centr. Hub & 3 & 2 & 0.949 & 0.046
 & 9.095 & 9.105 & 0.110 \\
Periph. Area & 1 & 1 & 0.950 & 0.217
 & 7.420 & 7.500 & 1.078 \\
Periph. Area & 3 & 1 & 0.947 & 0.252
 & 4.620 & 4.730 & 2.381 \\
\midrule
\multicolumn{8}{l}{\textit{Results by CLITO}}\\
\multicolumn{8}{l}{\textit{Covered}}\\
Centr. Hub & 0 & 1109 & 0.986 & 0.159
 & 8.376 & 8.422 & 0.551 \\
Centr. Hub & 1 & 27 & 0.989 & 0.112
 & 9.111 & 9.114 & 0.037 \\
Periph. Area & 0 & 220 & 0.986 & 0.178
 & 5.671 & 5.725 & 0.956 \\
Periph. Area & 1 & 4 & 0.995 & 0.118
 & 6.293 & 6.327 & 0.556 \\
\multicolumn{8}{l}{\textit{Not covered}}\\
Centr. Hub & 0 & 20 & 0.947 & 0.160
 & 6.873 & 6.923 & 0.720 \\
Centr. Hub & 1 & 1 & 0.949 & 0.037
 & 14.960 & 14.970 & 0.067 \\
Periph. Area & 0 & 2 & 0.948 & 0.235
 & 6.020 & 6.115 & 1.578 \\
\midrule
\multicolumn{8}{l}{\textit{Results by DEGURB}}\\
\multicolumn{8}{l}{\textit{Covered}}\\
Centr. Hub & 1 & 18 & 0.985 & 0.089
 & 14.414 & 14.419 & 0.031 \\
Centr. Hub & 2 & 490 & 0.986 & 0.160
 & 9.242 & 9.286 & 0.480 \\
Centr. Hub & 3 & 628 & 0.986 & 0.159
 & 7.559 & 7.606 & 0.620 \\
Periph. Area & 2 & 9 & 0.986 & 0.067
 & 8.437 & 8.448 & 0.132 \\
Periph. Area & 3 & 215 & 0.986 & 0.181
 & 5.567 & 5.623 & 1.000 \\
\multicolumn{8}{l}{\textit{Not covered}}\\
Centr. Hub & 1 & 1 & 0.949 & 0.037
 & 14.960 & 14.970 & 0.067 \\
Centr. Hub & 2 & 7 & 0.947 & 0.227
 & 9.231 & 9.319 & 0.944 \\
Centr. Hub & 3 & 13 & 0.946 & 0.125
 & 5.603 & 5.632 & 0.522 \\
Periph. Area & 3 & 2 & 0.948 & 0.235
 & 6.020 & 6.115 & 1.578 \\
\bottomrule
\end{tabular}
\end{table}
\begin{table}[ht]
\centering
\tiny
\captionsetup{
  width=\textwidth,
  justification=justified,
  singlelinecheck=false
}
\caption{Continuous Ising: MPI indices by territorial attribute classes}
\label{tab:c1_mpi_all}
\begin{tabular}{l r r r r r r r r}
\toprule
\textit{type} & \textit{class} & $y_{ref}$
& \textit{MPI1} & \textit{MPI2} & \textit{MPI3}
& \textit{MPI4} & \textit{MPI5} & \textit{MPI6} \\
\midrule
\multicolumn{9}{l}{\textit{Results by ALT}}\\
\multicolumn{9}{l}{\textit{Covered}}\\
Centr. Hub & 1 & 9.029 & 104.931 & 102.590
 & 107.559 & 107.258 & 102.064 & 99.654 \\
Centr. Hub & 2 & 6.418 & 100.039 & 100.186
 & 107.104 & 106.899 & 98.036 & 99.743 \\
Centr. Hub & 3 & 5.767 & 103.372 & 99.168
 & 108.353 & 108.417 & 98.552 & 99.530 \\
Periph. Area & 1 & 8.928 & 99.944 & 99.787
 & 99.043 & 102.458 & 108.931 & 97.977 \\
Periph. Area & 2 & 6.341 & 90.559 & 99.322
 & 105.026 & 105.471 & 100.717 & 99.490 \\
Periph. Area & 3 & 5.422 & 97.929 & 98.825
 & 106.318 & 106.234 & 100.515 & 99.121 \\
\multicolumn{9}{l}{\textit{Not covered}}\\
Centr. Hub & 1 & 7.949 & 103.557 & 101.748
 & 107.212 & 107.821 & 101.887 & 99.102 \\
Centr. Hub & 2 & 3.230 & 83.567 & 94.218
 & 102.163 & 97.561 & 100.419 & 99.726 \\
Centr. Hub & 3 & 5.992 & 106.592 & 98.804
 & 108.554 & 109.354 & 100.466 & 100.323 \\
Periph. Area & 3 & 6.020 & 96.878 & 103.456
 & 106.101 & 103.682 & 108.297 & 97.825 \\
\midrule
\multicolumn{9}{l}{\textit{Results by POP}}\\
\multicolumn{9}{l}{\textit{Covered}}\\
Centr. Hub & 1 & 7.094 & 102.321 & 99.888
 & 106.962 & 107.124 & 98.786 & 100.458 \\
Centr. Hub & 2 & 9.644 & 106.395 & 103.875
 & 108.236 & 107.665 & 103.353 & 99.172 \\
Centr. Hub & 3 & 13.898 & 101.484 & 112.937
 & 108.654 & 104.609 & 117.294 & 88.915 \\
Periph. Area & 1 & 5.510 & 96.476 & 98.749
 & 105.822 & 105.951 & 100.677 & 99.193 \\
Periph. Area & 2 & 8.079 & 105.139 & 101.420
 & 106.368 & 106.446 & 103.697 & 98.139 \\
\multicolumn{9}{l}{\textit{Not covered}}\\
Centr. Hub & 1 & 5.774 & 101.167 & 97.544
 & 106.860 & 107.198 & 97.083 & 100.429 \\
Centr. Hub & 2 & 8.217 & 106.802 & 103.475
 & 108.028 & 108.827 & 104.709 & 99.851 \\
Centr. Hub & 3 & 14.960 & 95.837 & 110.192
 & 105.392 & 103.030 & 120.765 & 84.494 \\
Periph. Area & 1 & 7.420 & 101.329 & 94.254
 & 102.805 & 103.691 & 90.839 & 97.591 \\
Periph. Area & 2 & 4.620 & 92.427 & 112.657
 & 109.398 & 103.672 & 125.754 & 98.059 \\
\midrule
\multicolumn{9}{l}{\textit{Results by SUP}}\\
\multicolumn{9}{l}{\textit{Covered}}\\
Centr. Hub & 1 & 7.352 & 106.257 & 102.120
 & 107.714 & 108.220 & 98.873 & 100.808 \\
Centr. Hub & 2 & 8.590 & 104.078 & 101.882
 & 107.545 & 107.270 & 101.386 & 99.689 \\
Centr. Hub & 3 & 9.125 & 99.821 & 102.391
 & 107.581 & 105.572 & 105.807 & 96.279 \\
Periph. Area & 1 & 5.614 & 98.124 & 99.517
 & 106.239 & 105.954 & 96.380 & 98.265 \\
Periph. Area & 2 & 5.782 & 96.827 & 99.383
 & 105.761 & 105.943 & 101.529 & 99.396 \\
Periph. Area & 3 & 5.312 & 97.254 & 96.607
 & 105.996 & 106.174 & 101.356 & 98.588 \\
\multicolumn{9}{l}{\textit{Not covered}}\\
Centr. Hub & 1 & 5.710 & 110.398 & 101.658
 & 108.140 & 110.109 & 100.963 & 101.160 \\
Centr. Hub & 2 & 7.224 & 104.099 & 100.396
 & 107.604 & 108.285 & 100.468 & 100.078 \\
Centr. Hub & 3 & 9.095 & 89.702 & 102.205
 & 103.778 & 100.296 & 110.592 & 92.110 \\
Periph. Area & 1 & 7.420 & 101.329 & 94.254
 & 102.805 & 103.691 & 90.839 & 97.591 \\
Periph. Area & 3 & 4.620 & 92.427 & 112.657
 & 109.398 & 103.672 & 125.754 & 98.059 \\
\midrule
\multicolumn{9}{l}{\textit{Results by CLITO}}\\
\multicolumn{9}{l}{\textit{Covered}}\\
Centr. Hub & 0 & 8.376 & 104.283 & 101.944
 & 107.683 & 107.449 & 101.063 & 99.717 \\
Centr. Hub & 1 & 9.111 & 99.985 & 102.818
 & 103.330 & 102.307 & 108.129 & 97.031 \\
Periph. Area & 0 & 5.671 & 96.993 & 99.014
 & 106.010 & 106.036 & 100.736 & 99.170 \\
Periph. Area & 1 & 6.293 & 100.504 & 94.212
 & 97.543 & 103.125 & 108.782 & 96.513 \\
\multicolumn{9}{l}{\textit{Not covered}}\\
Centr. Hub & 0 & 6.873 & 103.703 & 100.213
 & 107.386 & 107.931 & 100.515 & 100.169 \\
Centr. Hub & 1 & 14.960 & 95.837 & 110.192
 & 105.392 & 103.030 & 120.765 & 84.494 \\
Periph. Area & 0 & 6.020 & 96.878 & 103.456
 & 106.101 & 103.682 & 108.297 & 97.825 \\
\midrule
\multicolumn{9}{l}{\textit{Results by DEGURB}}\\
\multicolumn{9}{l}{\textit{Covered}}\\
Centr. Hub & 1 & 14.414 & 101.309 & 113.013
 & 108.552 & 104.467 & 118.180 & 88.509 \\
Centr. Hub & 2 & 9.242 & 106.575 & 104.246
 & 108.105 & 107.828 & 103.383 & 99.334 \\
Centr. Hub & 3 & 7.559 & 102.396 & 99.868
 & 107.143 & 107.017 & 99.066 & 100.221 \\
Periph. Area & 2 & 8.437 & 104.235 & 103.311
 & 106.923 & 106.212 & 104.255 & 98.322 \\
Periph. Area & 3 & 5.567 & 96.755 & 98.745
 & 105.814 & 105.974 & 100.738 & 99.156 \\
\multicolumn{9}{l}{\textit{Not covered}}\\
Centr. Hub & 1 & 14.960 & 95.837 & 110.192
 & 105.392 & 103.030 & 120.765 & 84.494 \\
Centr. Hub & 2 & 9.231 & 107.099 & 101.618
 & 107.931 & 108.624 & 105.567 & 99.636 \\
Centr. Hub & 3 & 5.603 & 101.874 & 99.456
 & 107.092 & 107.558 & 97.794 & 100.456 \\
Periph. Area & 3 & 6.020 & 96.878 & 103.456
 & 106.101 & 103.682 & 108.297 & 97.825 \\
\bottomrule
\end{tabular}
\end{table}
\begin{table}[ht]
\centering
\tiny
\captionsetup{
  width=\textwidth,
  justification=justified,
  singlelinecheck=false
}
\caption{Langevin dynamics: summary results by territorial attributes}
\label{tab:c2_summary_all}
\begin{tabular}{l r r r r r r r}
\toprule
\textit{type} & \textit{class} & \textit{n}
& \textit{coverage} & \textit{adaptivity}
& $y_{ref}$ & $y_{est}$ & $\delta$ (\%) \\
\midrule
\multicolumn{8}{l}{\textit{Results by ALT}}\\
\multicolumn{8}{l}{\textit{Covered}}\\
Centr. Hub & 1 & 886 & 0.985 & 0.243
 & 8.994 & 8.875 & -1.318 \\
Centr. Hub & 2 & 137 & 0.984 & 0.095
 & 6.434 & 6.389 & -0.702 \\
Centr. Hub & 3 & 115 & 0.987 & 0.170
 & 5.777 & 5.691 & -1.498 \\
Periph. Area & 1 & 8 & 0.978 & 0.061
 & 9.849 & 9.824 & -0.254 \\
Periph. Area & 2 & 29 & 0.985 & 0.094
 & 6.341 & 6.300 & -0.653 \\
Periph. Area & 3 & 185 & 0.986 & 0.122
 & 5.444 & 5.380 & -1.165 \\
\multicolumn{8}{l}{\textit{Not covered}}\\
Centr. Hub & 1 & 16 & 0.949 & 0.146
 & 9.975 & 9.917 & -0.583 \\
Centr. Hub & 2 & 3 & 0.949 & 0.084
 & 4.603 & 4.560 & -0.941 \\
Periph. Area & 1 & 1 & 0.943 & 0.017
 & 1.560 & 1.570 & 0.641 \\
Periph. Area & 3 & 3 & 0.947 & 0.050
 & 4.497 & 4.497 & 0.000 \\
\midrule
\multicolumn{8}{l}{\textit{Results by POP}}\\
\multicolumn{8}{l}{\textit{Covered}}\\
Centr. Hub & 1 & 593 & 0.985 & 0.159
 & 7.063 & 6.985 & -1.106 \\
Centr. Hub & 2 & 526 & 0.985 & 0.288
 & 9.620 & 9.479 & -1.466 \\
Centr. Hub & 3 & 19 & 0.981 & 0.080
 & 13.976 & 13.941 & -0.252 \\
Periph. Area & 1 & 207 & 0.986 & 0.122
 & 5.556 & 5.493 & -1.128 \\
Periph. Area & 2 & 15 & 0.983 & 0.036
 & 7.984 & 7.973 & -0.134 \\
\multicolumn{8}{l}{\textit{Not covered}}\\
Centr. Hub & 1 & 10 & 0.948 & 0.132
 & 7.452 & 7.407 & -0.604 \\
Centr. Hub & 2 & 6 & 0.949 & 0.190
 & 9.603 & 9.513 & -0.937 \\
Centr. Hub & 3 & 3 & 0.948 & 0.043
 & 13.757 & 13.733 & -0.170 \\
Periph. Area & 1 & 3 & 0.945 & 0.031
 & 3.003 & 2.993 & -0.333 \\
Periph. Area & 2 & 1 & 0.947 & 0.075
 & 6.040 & 6.080 & 0.662 \\
\midrule
\multicolumn{8}{l}{\textit{Results by SUP}}\\
\multicolumn{8}{l}{\textit{Covered}}\\
Centr. Hub & 1 & 212 & 0.986 & 0.124
 & 7.358 & 7.306 & -0.711 \\
Centr. Hub & 2 & 844 & 0.985 & 0.253
 & 8.555 & 8.429 & -1.471 \\
Centr. Hub & 3 & 82 & 0.983 & 0.098
 & 8.955 & 8.908 & -0.530 \\
Periph. Area & 1 & 27 & 0.986 & 0.063
 & 5.762 & 5.740 & -0.386 \\
Periph. Area & 2 & 158 & 0.986 & 0.142
 & 5.809 & 5.733 & -1.312 \\
Periph. Area & 3 & 37 & 0.985 & 0.046
 & 5.309 & 5.296 & -0.249 \\
\multicolumn{8}{l}{\textit{Not covered}}\\
Centr. Hub & 1 & 6 & 0.948 & 0.070
 & 6.583 & 6.585 & 0.025 \\
Centr. Hub & 2 & 10 & 0.949 & 0.204
 & 9.264 & 9.164 & -1.079 \\
Centr. Hub & 3 & 3 & 0.948 & 0.043
 & 13.757 & 13.733 & -0.170 \\
Periph. Area & 1 & 1 & 0.944 & 0.004
 & 3.440 & 3.440 & 0.000 \\
Periph. Area & 2 & 1 & 0.943 & 0.017
 & 1.560 & 1.570 & 0.641 \\
Periph. Area & 3 & 2 & 0.948 & 0.074
 & 5.025 & 5.025 & 0.000 \\
\midrule
\multicolumn{8}{l}{\textit{Results by CLITO}}\\
\multicolumn{8}{l}{\textit{Covered}}\\
Centr. Hub & 0 & 1111 & 0.985 & 0.219
 & 8.341 & 8.233 & -1.298 \\
Centr. Hub & 1 & 27 & 0.981 & 0.154
 & 9.172 & 9.138 & -0.372 \\
Periph. Area & 0 & 219 & 0.986 & 0.117
 & 5.690 & 5.630 & -1.053 \\
Periph. Area & 1 & 3 & 0.981 & 0.018
 & 7.870 & 7.867 & -0.042 \\
\multicolumn{8}{l}{\textit{Not covered}}\\
Centr. Hub & 0 & 18 & 0.949 & 0.139
 & 8.894 & 8.838 & -0.637 \\
Centr. Hub & 1 & 1 & 0.947 & 0.083
 & 13.310 & 13.270 & -0.301 \\
Periph. Area & 0 & 3 & 0.947 & 0.050
 & 4.497 & 4.497 & 0.000 \\
Periph. Area & 1 & 1 & 0.943 & 0.017
 & 1.560 & 1.570 & 0.641 \\
\midrule
\multicolumn{8}{l}{\textit{Results by DEGURB}}\\
\multicolumn{8}{l}{\textit{Covered}}\\
Centr. Hub & 1 & 16 & 0.984 & 0.091
 & 14.572 & 14.530 & -0.287 \\
Centr. Hub & 2 & 489 & 0.984 & 0.276
 & 9.243 & 9.106 & -1.477 \\
Centr. Hub & 3 & 633 & 0.985 & 0.176
 & 7.522 & 7.437 & -1.128 \\
Periph. Area & 2 & 9 & 0.986 & 0.070
 & 8.437 & 8.427 & -0.119 \\
Periph. Area & 3 & 213 & 0.986 & 0.118
 & 5.605 & 5.544 & -1.092 \\
\multicolumn{8}{l}{\textit{Not covered}}\\
Centr. Hub & 1 & 3 & 0.948 & 0.043
 & 13.757 & 13.733 & -0.170 \\
Centr. Hub & 2 & 8 & 0.949 & 0.154
 & 9.161 & 9.088 & -0.805 \\
Centr. Hub & 3 & 8 & 0.949 & 0.154
 & 7.356 & 7.306 & -0.680 \\
Periph. Area & 3 & 4 & 0.946 & 0.042
 & 3.763 & 3.765 & 0.066 \\
\bottomrule
\end{tabular}
\end{table}
\begin{table}[ht]
\centering
\tiny
\captionsetup{
  width=\textwidth,
  justification=justified,
  singlelinecheck=false
}
\caption{Langevin dynamics: MPI indices by territorial attribute classes}
\label{tab:c2_mpi_all}
\begin{tabular}{l r r r r r r r r}
\toprule
\textit{type} & \textit{class} & $y_{ref}$
& \textit{MPI1} & \textit{MPI2} & \textit{MPI3}
& \textit{MPI4} & \textit{MPI5} & \textit{MPI6} \\
\midrule
\multicolumn{9}{l}{\textit{Results by ALT}}\\
\multicolumn{9}{l}{\textit{Covered}}\\
Centr. Hub & 1 & 8.994 & 104.869 & 102.515
 & 107.551 & 107.269 & 102.007 & 99.676 \\
Centr. Hub & 2 & 6.434 & 99.758 & 100.266
 & 107.075 & 106.844 & 98.032 & 99.736 \\
Centr. Hub & 3 & 5.777 & 103.512 & 99.153
 & 108.361 & 108.458 & 98.635 & 99.564 \\
Periph. Area & 1 & 9.849 & 100.111 & 102.003
 & 100.702 & 101.567 & 109.085 & 98.205 \\
Periph. Area & 2 & 6.341 & 90.559 & 99.322
 & 105.026 & 105.471 & 100.717 & 99.490 \\
Periph. Area & 3 & 5.444 & 97.994 & 98.840
 & 106.379 & 106.291 & 100.733 & 99.128 \\
\multicolumn{9}{l}{\textit{Not covered}}\\
Centr. Hub & 1 & 9.975 & 107.038 & 105.948
 & 107.643 & 107.181 & 105.057 & 97.913 \\
Centr. Hub & 2 & 4.603 & 107.384 & 94.541
 & 106.743 & 106.269 & 98.984 & 100.037 \\
Periph. Area & 1 & 1.560 & 98.609 & 82.062
 & 85.772 & 109.584 & 107.697 & 96.153 \\
Periph. Area & 3 & 4.497 & 93.218 & 101.026
 & 102.400 & 101.026 & 92.275 & 97.782 \\
\midrule
\multicolumn{9}{l}{\textit{Results by POP}}\\
\multicolumn{9}{l}{\textit{Covered}}\\
Centr. Hub & 1 & 7.063 & 102.206 & 99.846
 & 106.952 & 107.116 & 98.767 & 100.453 \\
Centr. Hub & 2 & 9.620 & 106.395 & 103.833
 & 108.235 & 107.686 & 103.333 & 99.183 \\
Centr. Hub & 3 & 13.976 & 100.694 & 112.780
 & 108.784 & 104.632 & 117.346 & 88.849 \\
Periph. Area & 1 & 5.556 & 96.569 & 98.801
 & 105.946 & 106.003 & 100.701 & 99.210 \\
Periph. Area & 2 & 7.984 & 104.414 & 101.988
 & 106.717 & 106.157 & 105.589 & 98.209 \\
\multicolumn{9}{l}{\textit{Not covered}}\\
Centr. Hub & 1 & 7.452 & 107.881 & 99.809
 & 107.442 & 107.661 & 98.067 & 100.736 \\
Centr. Hub & 2 & 9.603 & 107.023 & 106.941
 & 107.980 & 107.546 & 107.139 & 99.295 \\
Centr. Hub & 3 & 13.757 & 104.603 & 113.016
 & 106.740 & 103.936 & 118.120 & 87.864 \\
Periph. Area & 1 & 3.003 & 91.656 & 93.672
 & 96.267 & 101.549 & 95.716 & 97.496 \\
Periph. Area & 2 & 6.040 & 103.294 & 104.123
 & 104.171 & 108.016 & 97.373 & 97.012 \\
\midrule
\multicolumn{9}{l}{\textit{Results by SUP}}\\
\multicolumn{9}{l}{\textit{Covered}}\\
Centr. Hub & 1 & 7.358 & 106.188 & 102.049
 & 107.708 & 108.226 & 98.972 & 100.797 \\
Centr. Hub & 2 & 8.555 & 104.055 & 101.859
 & 107.548 & 107.293 & 101.328 & 99.699 \\
Centr. Hub & 3 & 8.955 & 99.399 & 101.998
 & 107.519 & 105.503 & 105.474 & 96.486 \\
Periph. Area & 1 & 5.762 & 98.549 & 99.210
 & 106.228 & 105.980 & 96.202 & 98.225 \\
Periph. Area & 2 & 5.809 & 96.816 & 99.492
 & 105.887 & 105.920 & 101.490 & 99.416 \\
Periph. Area & 3 & 5.309 & 97.250 & 96.843
 & 106.301 & 106.438 & 102.597 & 98.641 \\
\multicolumn{9}{l}{\textit{Not covered}}\\
Centr. Hub & 1 & 6.583 & 110.048 & 104.456
 & 108.072 & 108.634 & 96.062 & 101.304 \\
Centr. Hub & 2 & 9.264 & 106.066 & 101.300
 & 107.387 & 107.008 & 104.713 & 99.531 \\
Centr. Hub & 3 & 13.757 & 104.603 & 113.016
 & 106.740 & 103.936 & 118.120 & 87.864 \\
Periph. Area & 1 & 3.440 & 89.850 & 102.543
 & 103.091 & 102.983 & 95.651 & 98.670 \\
Periph. Area & 2 & 1.560 & 98.609 & 82.062
 & 85.772 & 109.584 & 107.697 & 96.153 \\
Periph. Area & 3 & 5.025 & 94.902 & 100.268
 & 102.054 & 100.048 & 90.587 & 97.338 \\
\midrule
\multicolumn{9}{l}{\textit{Results by CLITO}}\\
\multicolumn{9}{l}{\textit{Covered}}\\
Centr. Hub & 0 & 8.341 & 104.224 & 101.885
 & 107.676 & 107.458 & 101.012 & 99.739 \\
Centr. Hub & 1 & 9.172 & 99.710 & 102.701
 & 103.458 & 102.401 & 108.427 & 96.917 \\
Periph. Area & 0 & 5.690 & 97.044 & 99.027
 & 106.060 & 106.083 & 100.921 & 99.177 \\
Periph. Area & 1 & 7.870 & 101.136 & 98.261
 & 101.467 & 100.972 & 109.143 & 96.632 \\
\multicolumn{9}{l}{\textit{Not covered}}\\
Centr. Hub & 0 & 8.894 & 107.304 & 103.635
 & 107.810 & 107.401 & 103.619 & 98.841 \\
Centr. Hub & 1 & 13.310 & 103.277 & 113.352
 & 101.944 & 100.480 & 112.721 & 87.586 \\
Periph. Area & 0 & 4.497 & 93.218 & 101.026
 & 102.400 & 101.026 & 92.275 & 97.782 \\
Periph. Area & 1 & 1.560 & 98.609 & 82.062
 & 85.772 & 109.584 & 107.697 & 96.153 \\
\midrule
\multicolumn{9}{l}{\textit{Results by DEGURB}}\\
\multicolumn{9}{l}{\textit{Covered}}\\
Centr. Hub & 1 & 14.572 & 100.349 & 112.836
 & 108.695 & 104.477 & 118.353 & 88.379 \\
Centr. Hub & 2 & 9.243 & 106.560 & 104.183
 & 108.105 & 107.832 & 103.410 & 99.330 \\
Centr. Hub & 3 & 7.522 & 102.325 & 99.869
 & 107.139 & 107.029 & 99.037 & 100.222 \\
Periph. Area & 2 & 8.437 & 104.235 & 103.311
 & 106.923 & 106.212 & 104.255 & 98.322 \\
Periph. Area & 3 & 5.605 & 96.797 & 98.835
 & 105.959 & 106.005 & 100.895 & 99.177 \\
\multicolumn{9}{l}{\textit{Not covered}}\\
Centr. Hub & 1 & 13.757 & 104.603 & 113.016
 & 106.740 & 103.936 & 118.120 & 87.864 \\
Centr. Hub & 2 & 9.161 & 107.920 & 105.809
 & 107.931 & 108.297 & 103.676 & 99.829 \\
Centr. Hub & 3 & 7.356 & 107.199 & 99.158
 & 107.357 & 106.939 & 99.262 & 100.562 \\
Periph. Area & 3 & 3.763 & 94.566 & 96.285
 & 98.243 & 103.166 & 96.130 & 97.375 \\
\bottomrule
\end{tabular}
\end{table}
Across both procedures, the covered groups have mean empirical
coverage close to $0.98$--$0.99$. The direction of the deviations
in their municipal means differs between procedures. All covered
groups reported for the Continuous Ising model have positive
relative differences, whereas the corresponding Langevin groups
have negative differences. The former therefore have simulated
group means above their observed references, while the latter
have simulated means below them. These group summaries do not
imply that every municipality within a group has the same
sign of deviation.

The differences remain modest at the group level. Among covered
groups, the relative differences range from $0.025\%$ to
$1.245\%$ for the Continuous Ising model and from
$-1.498\%$ to $-0.042\%$ for Langevin dynamics.
Coverage below the nominal level can nevertheless occur in
groups with very small differences between observed and
simulated means. For example, the uncovered large-population
group in the Continuous Ising model contains one municipality
with a relative difference of $0.067\%$ and coverage of $0.949$.
Agreement between municipal means and empirical interval
coverage therefore describe distinct aspects of the results. 
The ordering of interval widths depends on the territorial
classification and simulation procedure. Among covered small
municipalities (POP class 1), mean adaptivity is $0.161$
for the first municipality type and $0.184$ for the second
under the Continuous Ising model. Under Langevin dynamics,
the corresponding values are $0.159$ and $0.122$.
For covered medium-population municipalities of the first
type, Langevin dynamics produces a mean adaptivity of $0.288$,
compared with $0.158$ for the Continuous Ising model.
These results preclude a uniform ordering of interval widths
either between municipality types or between procedures. 
The socio-economic profiles show both recurring contrasts
and differences specific to individual territorial classes.
Among covered lowland municipalities (ALT class 1), the first
type has higher mean MPI3 and MPI4 values than the second
under both procedures. In the Continuous Ising results,
these values are $107.559$ versus $99.043$ for MPI3 and
$107.258$ versus $102.458$ for MPI4. The ordering differs
for MPI5, whose means are $102.064$ and $108.931$.
The corresponding Langevin groups exhibit the same directions
of contrast. Economic well-being, employment level, and
territorial attractiveness therefore do not define a single
common ordering across municipality types. 
MPI6 provides an additional dimension of these profiles,
distinct from demographic structure as represented by MPI1.
For covered mountain municipalities (ALT class 3), the
Continuous Ising results give mean MPI1 values of $103.372$
and $97.929$ for the two municipality types, whereas mean
MPI6 values are $99.530$ and $99.121$.
The contrast in demographic structure is thus accompanied
by a much smaller difference in population dynamism for
these groups. The separate indices allow such differences
between socio-economic dimensions to remain visible. 
Groups classified as not covered contain few municipalities,
with several cells representing a single municipality.
Their averages describe those specific units and should not
be interpreted as stable characteristics of broader territorial
categories. Moreover, coverage membership is procedure-specific,
so comparisons between the two sets of group summaries can
involve different municipalities even when the group labels
are identical. 
The descriptive comparisons associate socio-economic profiles
with differences in simulated municipal means, conformal
interval widths, and empirical coverage. Interval width
describes the extent of the calibrated range, whereas coverage
records its frequency of inclusion of the observed municipal
reference. Their joint examination characterises territorial
differences under the simulation protocol, without establishing
that economic conditions causally determine those differences.


\section{Discussion}\label{sec6}

The framework uses Statistical Mechanics to investigate local
territorial variation under register-derived interactions and
socio-economic drivers. Its purpose is exploratory: to examine
the observed configuration through nearby alternatives rather
than maximise predictive accuracy or identify a global optimum.

\subsection{Methodological contribution}

The central methodological choice is to treat the observed
territorial configuration as an empirical starting condition.
With the interaction graph and external field held fixed,
stochastic exploration asks whether the specified forces
generate energetically favourable alternatives that preserve
the observed organisation, and where departures occur.
Relating these departures to municipal characteristics connects
the energetic evaluation of configurations with substantive
questions about territorial organisation. 
Conformal interval widths and empirical coverage give this
assessment a municipal dimension, identifying cases where
the adopted representation merits closer examination.
These diagnostics reflect the combined simulation and
calibration procedure; discrepancies cannot be attributed
to the external field alone. 
The two stochastic strategies examine this compatibility
through distinct energy specifications and transition
mechanisms. Shared patterns indicate consistency across
implementations using common territorial inputs, while
differences reveal the dependence of simulated variation
on the chosen procedure.

\subsection{Interpretation of the empirical results}

Close agreement with observed municipal averages coexists
with different distributions of calibrated interval widths.
Langevin dynamics has a lower median but a higher mean width
than the Continuous Ising model, reflecting its more pronounced
upper tail. Neither procedure therefore produces uniformly
narrower intervals, and similar average coverage leaves
municipal differences relevant to territorial comparisons. 
The comparisons across strata challenge a simple
centre--periphery interpretation. Favourable economic and
occupational profiles do not imply uniformly lower adaptivity,
nor do their contrasts necessarily coincide with those in
territorial attractiveness or population dynamism.
Combinations of municipal attributes provide a more informative
basis for interpretation than centrality alone. 
The direction of simulated deviations also distinguishes the
procedures. Across the covered groups reported in Section~6,
Continuous Ising estimates exceed observed group means,
whereas Langevin estimates fall below them. Although small,
these systematic differences show that agreement in overall
levels does not imply identical responses to shared inputs.
Their interpretation requires consideration of the energy
specifications and update rules, rather than attribution
to an individual socio-economic driver. 
Agreement between procedures also reflects common inputs
and assumptions. Separately varying the external field,
interaction graph, and simulation settings would help assess
which associations persist and distinguish their contributions.
Comparisons across additional chains would clarify dependence
on the retained ensemble.

\subsection{Implications for Official Statistics}

Register-derived indicators can support explicit hypotheses
about relationships between municipalities, extending their
use beyond descriptions of individual units. The framework
provides a setting for exploring these hypotheses and assessing
the adequacy of territorial groupings: differences within
classes may motivate finer distinctions, while similarities
across classes may suggest alternative groupings. 
Such analyses can inform the interpretation and communication
of territorial diversity using existing statistical sources.
Broader applications require evidence across indicators,
territorial scales, and plausible specifications, supported
by sensitivity analysis and transparent documentation.
The present study provides an initial empirical basis for
assessing that practical value.


\section{Conclusions}\label{sec7}

This study proposes an energy-based framework for investigating
the compatibility of observed territorial configurations with
interactions and socio-economic drivers derived from statistical
registers. The observed municipal values define the initial
state and fixed empirical reference; the interaction structure
and external field remain unchanged during exploration. 
Applied to resident foreign population shares, both stochastic
procedures generate lower-energy configurations while retaining
municipal averages close to observed values. Their differing
local variations and associations with socio-economic profiles
reveal territorial heterogeneity that cannot be reduced to
a uniform centre--periphery contrast. 
Conformal interval widths and empirical coverage locate
municipalities warranting closer examination under the combined
modelling and calibration framework. They support questions
about the adequacy of the territorial representation;
isolating individual drivers requires targeted comparisons
that vary the field or interaction structure. 
The contribution to Official Statistics is an exploratory link
between observed territorial patterns, nearby simulated
alternatives, and hypotheses about their structural organisation.
Future work can examine other continuous indicators, alternative
interaction structures, and sensitivity to simulation settings.
Observations from multiple periods would enable extensions
to changes in territorial organisation over time.


\section*{Declarations}

\begin{itemize}
  \item \textit{Funding} \\  
    The author declares that this research received no external funding.
	\\
  \item \textit{Conflict of interest/Competing interests} \\  
    The author declares no conflicts of interest regarding the content of this manuscript.
  \\
   \item \textit{Ethics approval and consent to participate} \\  
    Not applicable.
    \\ 
  \item \textit{Consent for publication} \\  
   Not applicable.
    \\  
  \item \textit{Data availability} \\  
    The data supporting the findings of this study are available
    from the corresponding author upon reasonable request.
     \\ 
  \item \textit{Materials availability} \\  
    Not applicable. 
    \\
	\item \textit{Code availability} \\  
	The code supporting the findings 
	of this study is available upon 
	reasonable request. \\
\end{itemize}





\end{document}